\documentclass[twocolumn,preprintnumbers,aps,pra,superscriptaddress,amsmath,amssymb]{revtex4-1}

\usepackage{graphicx}
\usepackage{dcolumn}
\usepackage{bm}
\usepackage{bbm}
\usepackage{hyperref}
\usepackage{color}
\usepackage{multirow}

\begin{document}

\title{Wide-band, nanoscale magnetic resonance spectroscopy using quantum relaxation of a single spin in diamond}
	
\author{James D. A. Wood}
\affiliation{Centre for Quantum Computation and Communication Technology, School of Physics, The University of Melbourne, VIC 3010, Australia}

\author{David A. Broadway}
\affiliation{Centre for Quantum Computation and Communication Technology, School of Physics, The University of Melbourne, VIC 3010, Australia}

\author{Liam T. Hall}
\affiliation{School of Physics, The University of Melbourne, VIC 3010, Australia}

\author{Alastair Stacey}
\affiliation{Centre for Quantum Computation and Communication Technology, School of Physics, The University of Melbourne, VIC 3010, Australia}

\author{David A. Simpson}
\affiliation{School of Physics, The University of Melbourne, VIC 3010, Australia}
	
\author{Jean-Philippe Tetienne}\email{jtetienne@unimelb.edu.au}
\affiliation{Centre for Quantum Computation and Communication Technology, School of Physics, The University of Melbourne, VIC 3010, Australia}

\author{Lloyd C. L. Hollenberg}
\affiliation{Centre for Quantum Computation and Communication Technology, School of Physics, The University of Melbourne, VIC 3010, Australia}
\affiliation{School of Physics, The University of Melbourne, VIC 3010, Australia}

\date{\today}
	
\begin{abstract}
We demonstrate a wide-band all-optical method of nanoscale magnetic resonance (MR) spectroscopy under ambient conditions. Our method relies on cross-relaxation between a probe spin, the electronic spin of a nitrogen-vacancy centre in diamond, and target spins as the two systems are tuned into resonance. By optically monitoring the spin relaxation time ($T_1$) of the probe spin while varying the amplitude of an applied static magnetic field, a frequency spectrum of the target spin resonances, a $T_1$-MR spectrum, is obtained. As a proof of concept, we measure $T_1$-MR spectra of a small ensemble of $^{14}$N impurities surrounding the probe spin within the diamond, with each impurity comprising an electron spin 1/2 and a nuclear spin 1. The intrinsically large bandwidth of the technique and probe properties allows us to detect both electron spin transitions -- in the GHz range -- and nuclear spin transitions -- in the MHz range -- of the $^{14}$N spin targets. The measured frequencies are found to be in excellent agreement with theoretical expectations, and allow us to infer the hyperfine, quadrupole and gyromagnetic constants of the target spins. Analysis of the strength of the resonances obtained in the $T_1$-MR spectrum reveals that the electron spin transitions are probed via dipole interactions, while the nuclear spin resonances are dramatically enhanced by hyperfine coupling and an electron-mediated process. Finally, we investigate theoretically the possibility of performing $T_1$-MR spectroscopy on nuclear spins without hyperfine interaction and predict single-proton sensitivity using current technology. This work establishes $T_1$-MR as a simple yet powerful technique for nanoscale MR spectroscopy, with broadband capability and a projected sensitivity down to the single nuclear spin level. 
\end{abstract}	

\maketitle

\section{Introduction}

Magnetic resonance (MR) spectroscopic techniques are of primary importance in a variety of fields from physics to materials science, chemistry and biology. Electron paramagnetic resonance (EPR) spectroscopy enables characterisation of electronic systems in materials and systems containing unpaired electron spins, such as metal complexes and organic radicals; whereas nuclear magnetic resonance (NMR) spectroscopy probes nuclei with a non-zero spin, and is routinely used in chemical analysis of macromolecules. However, conventional EPR and NMR methods require macroscopic samples composed of millions of spins, impeding their use in the investigation of nanoscale materials and processes.
The nitrogen-vacancy (NV) centre in diamond (see \cite{Doherty2013} for a review) has been developed as a nanoscale magnetometer \cite{Rondin2014,Schirhagl2014} for the detection of static \cite{Taylor2008,Balasubramanian2008}, oscillating \cite{Maze2008} and randomly fluctuating fields \cite{Cole2009,Hall2009,Laraoui2010}. Recently, pathways towards nanoscale MR spectroscopy using the NV centre have been proposed and demonstrated \cite{Laraoui2012,Mamin2012,Mamin2013,Staudacher2013,Loretz2014,Muller2014,DeVience2015,Hall2015}. These techniques broadly fall into two classes: measurements based on the NV dephasing time ($T_2$) and measurement of the NV relaxation time ($T_1$). The first of these uses the NV centre to sense the oscillating field produced by sample spins non-resonantly coupled to the NV spin. In these techniques, the dephasing rate ($1/T_2$) of the NV spin is monitored in response to either continuous driving of the target spins using a resonant microwave field \cite{Laraoui2012,Mamin2012,Mamin2013}; or by applying a frequency-selective dynamical-decoupling sequence to the NV probe spin in order to detect the Larmor precession of the target spins \cite{Staudacher2013,Loretz2014,Muller2014,DeVience2015}. In both cases, it is possible to extract spectral information about the target environment, with sensitivity and spectral resolution governed by the NV centre's intrinsic dephasing rate. These dephasing-based techniques however have limitations. Pulsed magnetic resonance techniques such as double electron electron resonance (DEER) and electron-nuclear double resonance (ENDOR) require the ability to magnetically drive the target spins, which poses a challenge for spins with short lifetimes or small gyromagnetic ratios \cite{Mamin2012}. NMR spectroscopy via measurement of the Larmor precession field requires the application of complex microwave pulse sequences. These are highly susceptible to pulsing errors and contain harmonic resonances away from the central interrogation frequency \cite{Loretz2015}. Moreover, all of these techniques have their interrogation times limited by the probe's $T_2$ dephasing time. For external spin detection, requiring high sensitivity, the reduction in $T_2$ as NV centres approach the diamond surface is a significant drawback which limits the number of spins that can be detected \cite{Rosskopf2014,Myers2014}.

In this work we pursue the second type of nanoscale MR spectroscopy based on $T_1$ measurements. $T_1$-MR as a means of extracting the spectral distribution of the nanoscale environment was developed by Hall et al. and demonstrated with an ensemble of NV probes \cite{Hall2015}. The $T_1$-MR technique involves measuring the longitudinal relaxation rate ($1/T_1$) of the NV probe as a function of a controlled static background field \cite{Jarmola2012,VanderSar2015}, causing the probe's rate of relaxation to increase as it is brought into resonance with target spin transitions \cite{Hall2015}. Relaxometry-based $T_1$-MR has the potential to overcome a number of issues associated with $T_2$-based MR, by removing the need to pulse either the probe or target spin, as well as allowing the interrogation time to be extended from the dephasing timescale, $T_2$, out to the relaxation timescale, $T_1$, of the probe, which for near-surface NV spins can be up to three orders of magnitude longer than $T_2$ \cite{Rosskopf2014,Myers2014}. Here we extend this concept to a single spin probe, demonstrate the technique's broadband applicability, and uncover a new mechanism for detecting NMR transitions. Using $^{14}$N impurities within the diamond as a test system, we show that this method allows probing of both EPR and NMR transitions under ambient conditions while using a single spin probe, without requiring any microwave or radiofrequency driving. This allows us to extract information about the target system including hyperfine parameters, the nuclear quadrupole coupling parameter and the nuclear gyromagnetic ratio. In this experiment, the NMR transitions are occur via a two-step process involving hyperfine interaction and electron-electron interaction, which dramatically enhances the signal strength. In addition, we  investigate theoretically the technique's potential to detect NMR transitions of single nuclear spins external to the diamond. We predict that spectroscopy at the single proton level is achievable under realistic conditions. With the prospect of further improving the sensitivity through materials optimisation, our approach constitutes a promising alternative to $T_2$-based techniques towards single-molecule NMR spectroscopy and imaging.   

This article is organised as follows. We first describe the general principle of the technique employed in this work (Sec. \ref{sec:principle}). We then present experimental results of EPR and NMR spectroscopy of $^{14}$N impurities in diamond (Sec. \ref{sec:exp}). In Sec. \ref{sec:NMR}, we extrapolate these results and consider theoretically the detection of single nuclear spins. Further experimental details, as well as full theoretical methods, are given in Sec. \ref{sec:methods}. 

\section{Technique Overview} \label{sec:principle}

\begin{figure*}[t]
\begin{center}
\includegraphics[width=0.9\textwidth]{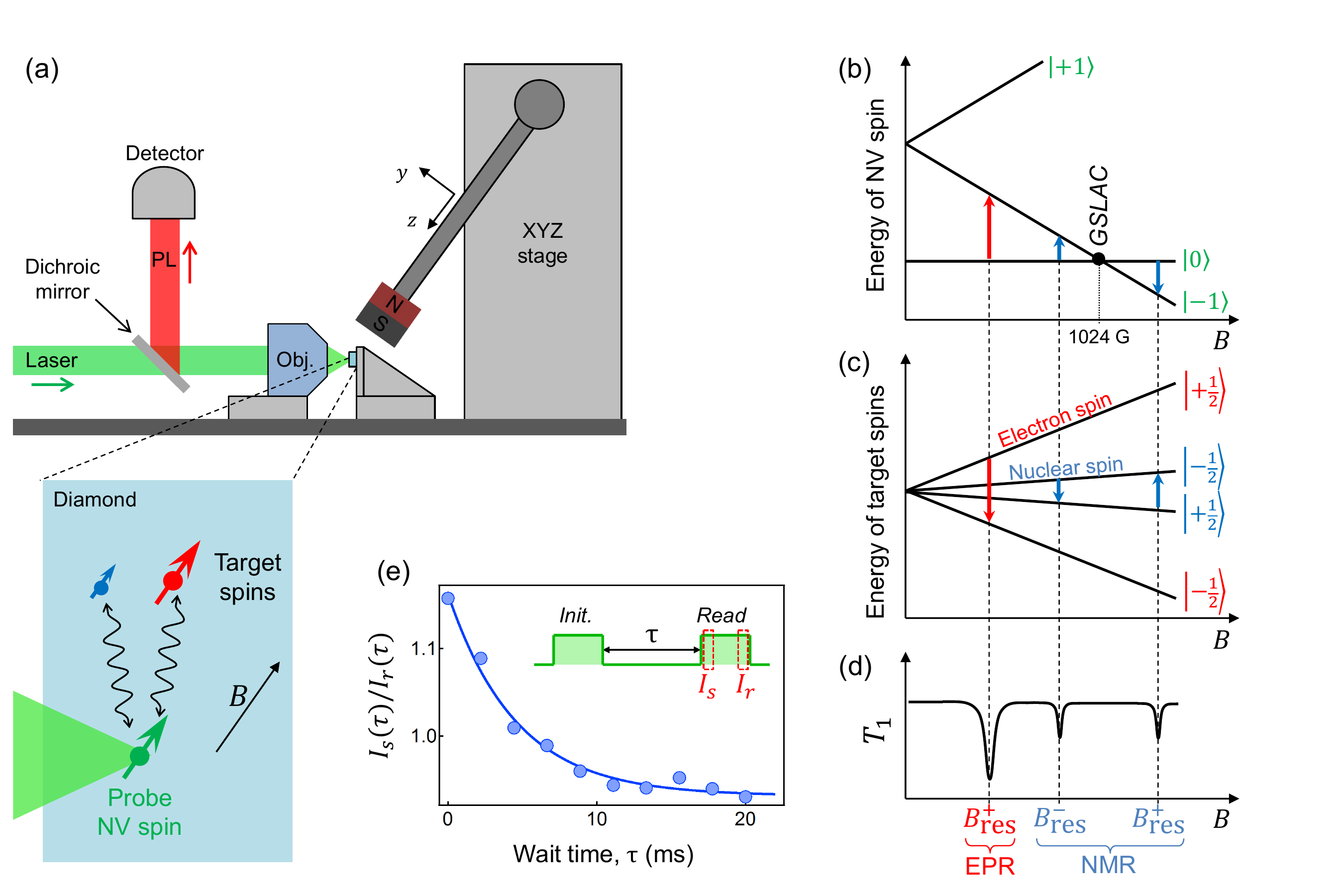}
\caption{(a) Schematic of the experimental setup with a purpose-built confocal microscope and a permanent magnet on a three-axis scanning stage. The electronic spin of an NV centre in diamond (green arrow) is used to probe environmental target spins (red and blue arrows) by optically measuring the relaxation time $T_1$ of the NV spin while scanning the strength $B$ of the applied magnetic field. (b) Energy levels of the NV spin as a function of $B$. The ground state level anti-crossing (GSLAC) occurs at $B_{\rm GSLAC}\approx1024$ G. (c) Energy levels of two spin-1/2 spins with different gyromagnetic ratios. These could be an electron and nuclear spin. (d) Schematic illustration of $T_1$-MR spectroscopy: the $T_1$ time of the NV spin is measured as a function of $B$ revealing cross-relaxation resonances whenever the transition frequency of the NV spin matches that of a target spin, as indicated by the red and blue arrows in (b) and (c). (e) Spin relaxation curve of a single NV centre off resonance. The inset shows the sequence of laser pulses, and the integration windows for the PL signal at the start ($I_s$) and end ($I_r$) of the readout pulse. The solid line is a single exponential fit, indicating a characteristic decay time $T_1=4.5\pm0.5$ ms for a typical probe spin.} 
\label{Fig1}
\end{center}
\end{figure*}

The schematic of the experimental setup is shown in Fig. \ref{Fig1}a. The probe spin is the electronic spin of an NV centre in diamond, which consists of a substitutional nitrogen adjacent to a vacancy \cite{Doherty2013}. The experimental setup includes a scanning confocal microscope allowing the single NV probe to be addressed with a 532 nm laser, with emitted red photoluminescence (PL) then measured by an avalanche photodiode detector. A permanent magnet is attached to a three-axis scanning stage in order to vary the direction and strength of the applied magnetic field (see further details about the setup in Sec. \ref{sec:methods:setup}). 

The NV centre's electronic ground state is a spin triplet with Hamiltonian
\begin{eqnarray} \label{eq:NVHam}
\frac{{\cal H}_{\rm NV}}{h} &=& D_{\rm NV}S_z^2 - \tilde{\gamma}_{\rm NV}B S_z
\end{eqnarray}
where $h$ is Planck's constant, $\tilde{\gamma}_{\rm NV} = -28.035(3)$ GHz/T is the gyromagnetic ratio of the NV electron spin \cite{NoteGamma,Felton2009}, and $D_{\rm NV}\approx2.87$ GHz is the crystal field splitting. Here $S_z$ refers to the spin-1 operator of the probe NV spin along the NV centre's symmetry axis, defined as the $z$ axis. The external magnetic field is aligned along $z$ with a strength $B$. The eigenstates of the probe (p) spin are denoted as $\vert m_S^{(\rm p)}\rangle$ where $m_S^{(\rm p)}$ refers to the spin projection along $z$. In zero magnetic field, the $\vert 0\rangle$ and $\vert \pm1\rangle$ states are split by a frequency $D_{\rm NV}$ (Fig. \ref{Fig1}b). Due to this zero-field splitting, environmental spin species can be brought into resonance with the NV's ground state transitions via the Zeeman effect. This is achieved by applying the appropriate static magnetic field along the NV symmetry axis (Fig. \ref{Fig1}c). Precisely, the NV spin transition $\vert 0\rangle\rightarrow\vert -1\rangle$ correspond to a frequency $\omega_{\rm NV}(B)=D_{\rm NV}+\tilde{\gamma}_{\rm NV}B$. The transition frequency of a target (electronic or nuclear) spin can be generally expressed as $\omega_{\rm t}(B)=D_{\rm t}+\tilde{\gamma}_{\rm t}B$ where $\tilde{\gamma}_{\rm t}$ is the gyromagnetic ratio of this particular spin and $D_{\rm t}$ is its intrinsic splitting, which may include zero-field splittings, quadrupolar interactions, hyperfine interactions, or interactions with the local environment such as dipole couplings or chemical shifts. The NV-target resonance condition is 
\begin{equation} \label{eq:rescond}
\left|\omega_{\rm NV}(B_{\rm res})\right|=\left|\omega_{\rm t}(B_{\rm res})\right|,
\end{equation}
from which one deduces the two resonant magnetic fields
\begin{equation} \label{eq:Bres}
B_{\rm res}^\pm=\left\vert\frac{D_{\rm NV}\pm D_{\rm t}}{\tilde{\gamma}_{\rm NV}\pm\tilde{\gamma}_{\rm t}}\right\vert.
\end{equation}
When this condition is fulfilled (i.e., $B=B_{\rm res}^\pm$), the NV and target spins can exchange energy through their mutual dipole-dipole interaction, which leads to an increased spin relaxation rate of both systems. Thus, monitoring the spin relaxation time $T_1$ of the NV centre whilst varying the axial field strength $B$, yields a $T_1$-MR spectrum exhibiting one or several resonant fields $B_{\rm res}$ (Fig. \ref{Fig1}d) corresponding to target spin transitions \cite{Hall2015}. For electronic spins, the resonances are typically centred about $B_{\rm res}^+\sim500$ G, corresponding to transition frequencies $\omega_{\rm t}(B_{\rm res}^+)\sim 1-2$ GHz. Owing to their much smaller gyromagnetic ratio, nuclear spins will interact with the NV at fields $B_{\rm res}^\pm\sim1000$ G, close to the $B=1024$ G ground state level anti-crossing of the NV centre (GSLAC, Fig. \ref{Fig1}b). This corresponds to transition frequencies ranging from a few MHz for an isolated nuclear spin with $D_{\rm t}=0$ (e.g., a proton $^1$H) up to $\sim100$ MHz in the presence of a hyperfine interaction with a nearby electron. By combining this with purely optical monitoring of the NV spin relaxation time $T_1$ \cite{Jarmola2012,Steinert2013,Tetienne2013,Kaufmann2013} (Fig. \ref{Fig1}e), an all-optical, broadband, nanoscale MR spectrometer may be realised. In Ref. \cite{Hall2015}, the technique was demonstrated using an ensemble of NV centres to obtain the EPR spectrum of $^{14}$N impurities in bulk diamond. In this work we extend the technique both to the single NV probe regime and from EPR to low frequency NMR spectroscopy. 

\section{Nanoscale $T_1$-spectroscopy using a single NV spin probe} \label{sec:exp}

To demonstrate the technique experimentally at the single NV level, we consider a target environment composed of $^{14}$N substitutional donor impurities residing in the same diamond crystal as the NV probe. These defects (usually referred to as P1 centres), when found in their uncharged state, comprise an electronic spin $S=1/2$ associated with an unpaired electron and a nuclear spin $I=1$ associated with the $^{14}$N nucleus \cite{Loubser1978,Smith1959,Cook1966,Hanson2008}. This configuration allows us to investigate both electronic (EPR) and nuclear (NMR) spin transitions, thereby demonstrating the broadband nature of the technique. Our sample is a type-Ib diamond from Element Six with a $^{14}$N concentration specified to be $<200$ ppm. The upper limit of 200 ppm corresponds to a median distance between an NV centre and the nearest P1 centre of $\approx1.7$ nm. Several individual NV centres were studied and gave consistent results. All measurements shown in the following have been obtained with the same NV centre, at room temperature. 

To measure the $T_1$ time of the NV spin, a 3-$\mu$s laser pulse is applied to initialise the spin into $\left\vert 0\right\rangle$, while a subsequent laser pulse reads out the spin state after a variable wait time $\tau$ (see inset in Fig. \ref{Fig1}e). This sequence is repeated many times while the time-resolved PL is monitored. The PL intensity immediately following the start of the pulse, the signal $I_s(\tau)$, is a measure of the population in the $\left\vert 0\right\rangle$ state \cite{Manson2006}, while the PL at the end of the pulse, $I_r(\tau)$, serves for normalisation purposes. The ratio $I_s(\tau)/I_r(\tau)$ therefore measures the decay out of $\left\vert 0\right\rangle$ after a time $\tau$. A typical relaxation curve is shown in Fig. \ref{Fig1}e, which is well described by a single exponential decay $\exp(-\tau/T_1)$. Away from any resonance with environmental spins, the characteristic decay time is typically $T_1\sim5$ ms, governed by two-phonon Orbach processes \cite{Jarmola2012}.

To vary the strength of the external magnetic field, the permanent magnet is scanned along the $z$ axis of the probe spin. The field direction is finely aligned along the NV centre's symmetry axis by exploiting the dependence of the PL intensity on the transverse magnetic field \cite{Epstein2005,Tetienne2012}. For each magnet position, an optically detected magnetic resonance (ODMR) spectrum of the NV spin is first recorded in order to determine the NV transition frequency $\omega_{\rm NV}$, from which the magnetic field amplitude $B$ is deduced. Next, the spin population decay after a fixed time $\tau$ is measured by repeating the laser pulse sequence $\approx10^6$ times, after which the magnet is moved to the next position. Further details regarding the acquisition procedure are given in Sec. \ref{sec:methods:procedure}.

\begin{figure*}[t]
\begin{center}
\includegraphics[width=0.8\textwidth]{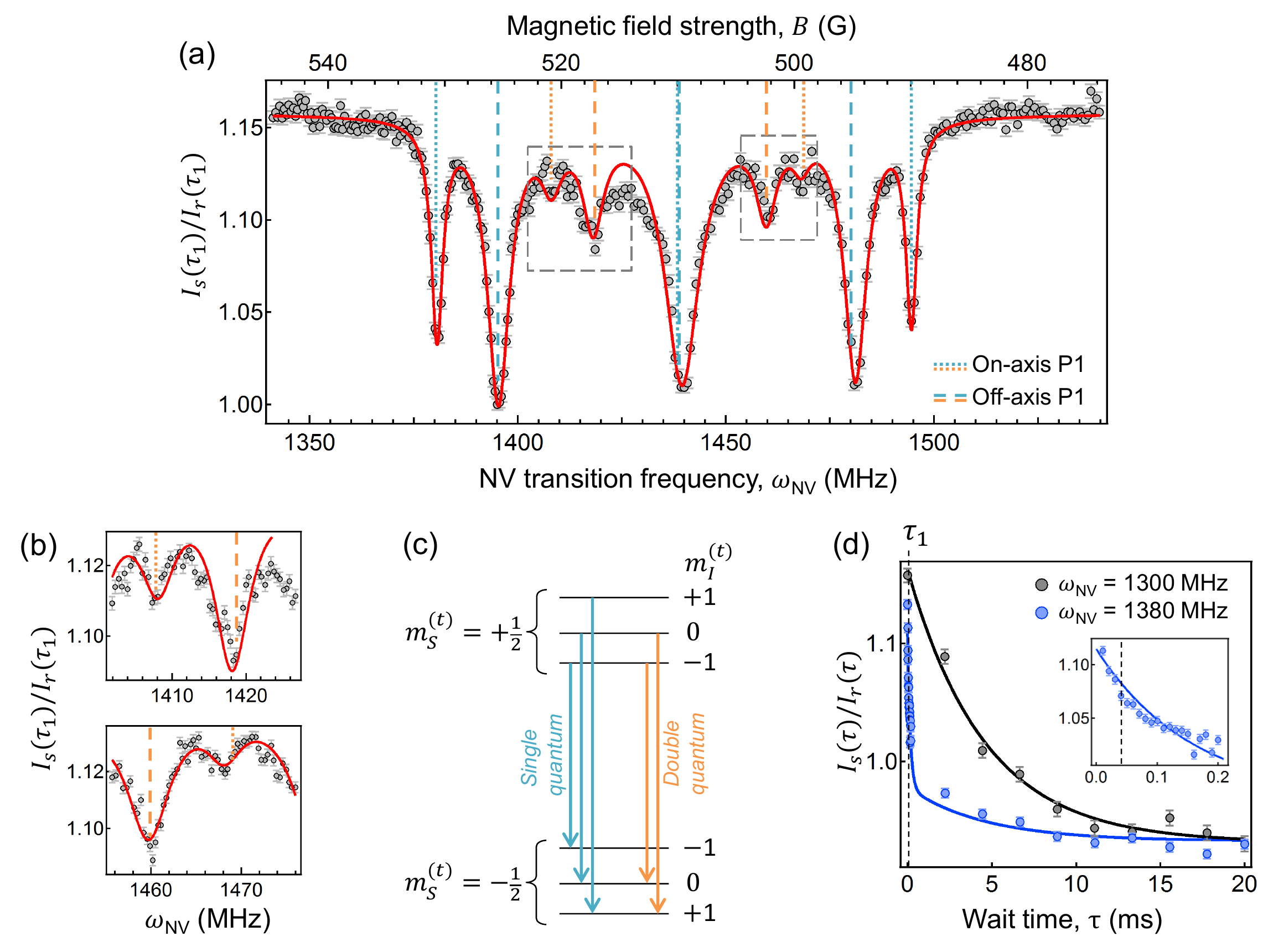}
\caption{(a) $T_1$-EPR spectrum of P1 centres in diamond obtained by measuring the population decay of a single NV spin after a wait time $\tau_1=40~\mu$s. The normalised PL signal $I_s(\tau_1)/I_r(\tau_1)$ is plotted against the NV transition frequency $\omega_{\rm NV}$, which is obtained from the ODMR spectrum. Also indicated is the corresponding magnetic field strength (top axis) obtained using Eq. (\ref{eq:NVfreq}). (b) High resolution spectra corresponding to the regions indicated by the dotted squares in (a). In (a,b), solid lines are data fitting to a sum of nine Lorentzian functions; Vertical lines indicate the theoretical frequencies for each allowed transition, dotted (dashed) lines corresponding to the on-axis (off-axis) P1 centres with colors as defined in (c). (c) Energy levels of a single P1 centre. The number $m_S^{(\rm t)}$ ($m_I^{(\rm t)}$) denotes the projection of the electron (nuclear) spin along the quantization axis fixed by the external magnetic field. Blue arrows represent the single-quantum transitions ($|\Delta m_S^{(\rm t)}|=1$, $\Delta m_I^{(\rm t)}=0$) while orange arrows represent the allowed double-quantum transitions ($|\Delta m_S^{(\rm t)}|=|\Delta m_I^{(\rm t)}|=1$).  (d) Full $T_1$ relaxation curve measured off resonance (black markers, $\omega_{\rm NV}=1300$ MHz) and on a P1 resonance (blue markers, $\omega_{\rm NV}=1380$ MHz). Inset: zoom-in of the on-resonance data for short evolution times (same units as in main graph). Solid lines are data fitting to Eq. (\ref{eq:T1curve}). Vertical dashed line indicates the probe time $\tau_1$ used in (a,b).} 
\label{Fig2}
\end{center}
\end{figure*}

\subsection{$T_1$-EPR spectroscopy of P1 centres}

In the first instance we measure the EPR spectrum of the P1 centres in the vicinity of a single NV centre probe, thus extending previous measurements to the single probe domain. To this end, the magnetic field is scanned in the range $B\approx 480-540$ G, corresponding to transition frequencies $\omega_{\rm NV}\approx1350-1530$ MHz. A probe evolution time of $\tau_1=40~\mu$s was used to monitor the NV spin relaxation rate and detect cross-relaxation events as $B$ is varied. This evolution time was chosen to optimise the sensitivity given the strength of the observed transitions. Figs. \ref{Fig2}a and \ref{Fig2}b show the normalised signal $I_s(\tau_1)/I_r(\tau_1)$ plotted against $\omega_{\rm NV}$ (bottom axis) and $B$ (top axis). These measurements confirm the presence of five main transitions seen in previous P1 studies \cite{Laraoui2012,deLange2012,Knowles2013}. These correspond to a change of the target (t) P1 electron spin projection ($m_S^{(\rm t)} =+\frac{1}{2}\rightarrow -\frac{1}{2}$) while the nuclear spin projection is conserved ($\Delta m_I^{(\rm t)}=0$), and are referred to as single-quantum transitions (Fig. \ref{Fig2}c). The presence of five transitions is caused by two effects. First, the hyperfine interaction induces a three-fold splitting associated with the three possible nuclear spin projections $m_I^{(\rm t)}=0,\pm1$. Second, there exists two families of P1 centres depending on whether the symmetry axis of the unpaired electron's orbital is parallel to the $[111]$ crystallographic axis (`on-axis'), which is also by convention the direction of the NV symmetry axis and of the applied magnetic field, or along one of the other three axes $[\bar{1}11]$, $[1\bar{1}1]$ and $[11\bar{1}]$ (`off-axis'). Consequently, the measured hyperfine splitting of the $m_I^{(\rm t)}=\pm1$ states have two different values for these two families of P1 centres, giving a total of five different transition frequencies. Note that each P1 centre switches between all four symmetry axes on a time scale of a few ms \cite{Hanson2008}, which is much shorter than our total measurement time. Therefore, even a single P1 centre will produce five resonance lines in the $T_1$-EPR spectrum, with the on-axis lines being three times weaker than the off-axis lines, since all four possible axes have an equal rate of occurrence. 

In addition to the five main spectral features, four weaker peaks are observed in the $T_1$-EPR spectrum (see zoom-in graphs in Fig. \ref{Fig2}b). These correspond to double-quantum transitions involving a flip of both the P1 electron spin ($m_S^{(\rm t)}=+\frac{1}{2}\rightarrow -\frac{1}{2}$) and nuclear spin ($\Delta m_I^{(\rm t)} = +1$) \cite{Hall2015}. The two allowed transitions are depicted in Fig. \ref{Fig2}c and are also split by the two families of P1 symmetry axis.

The transition frequencies are predicted by calculating the eigenvalues of the Hamiltonian of the P1 centre, ${\cal H}_{\rm P1}$, and solving the resonance condition (\ref{eq:rescond}). The spin Hamiltonian for an on-axis P1 centre is
\begin{eqnarray} \label{eq:HP1}
\frac{{\cal H}_{\rm P1}}{h} & = &-\tilde{\gamma}_eBS_z-\tilde{\gamma}_{N}BI_z+A_\parallel S_z I_z \nonumber \\
& & +A_\perp(S_x I_x+S_y I_y)+Q I_z^2 
\end{eqnarray}
where $\tilde{\gamma}_e= -28.024$ GHz/T and $\tilde{\gamma}_N=3.077$ MHz/T are the gyromagnetic ratios of the electron and of the $^{14}$N nucleus, respectively, $A_\parallel=113.98$ MHz and $A_\perp=81.34$ MHz are the axial and transverse hyperfine coupling parameters, and $Q=-3.97$ MHz is the nuclear quadrupole coupling parameter \cite{Cook1966}. Finally, $\vec{S}=(S_x,S_y,S_z)$ and $\vec{I}=(I_x,I_y,I_z)$ are the electron and nuclear spin operators of the target P1. The analytic expressions for the transition frequencies at resonance, $\omega_{\rm NV}(B_{\rm res})$, are given in Sec. \ref{sec:methods:P1}, and the resulting values are indicated in Table \ref{Table:EPR} and shown as vertical lines in Figs. \ref{Fig2}a and \ref{Fig2}b. They are found to be in excellent agreement with the experimental values.

\begin{table}[b]
\begin{tabular}{|c|c|c|c|}
\hline
\multirow{2}{*}{$m_I^{(\rm t)}$} & Symmetry & \multicolumn{2}{c|}{$\omega_{\rm NV}(B_{\rm res})$ (MHz)} \\
\cline{3-4}
& (on/off axis) & Theory & Experiment \\
\hline 
\hline
\multirow{2}{*}{$+1$} & on & 1380.1(1) & 1380.7(1) \\
\cline{2-4}
 & off &  1395.1(1) & 1395.4(1) \\
 \hline
\multirow{2}{*}{$0\rightarrow+1$} & on & 1407.9(1)  & 1408(1) \\
\cline{2-4}
 & off & 1418.7(1)  & 1418(1) \\
 \hline
\multirow{2}{*}{$0$} & on & 1438.3(1)  & \multirow{2}{*}{1439.5(1)} \\
\cline{2-3}
& off & 1439.2(1)  &  \\
\hline
\multirow{2}{*}{$-1\rightarrow0$} & off & 1459.9(1)  & 1460(1) \\
\cline{2-4}
 & on &  1469.0(1) & 1468(1) \\
 \hline
\multirow{2}{*}{$-1$} & off & 1480.2(1)   & 1481.2(1) \\
\cline{2-4}
 & on & 1494.2(1)  & 1494.6(1) \\
\hline
\end{tabular}
\caption{Summary of the theoretical and experimental EPR transition frequencies of P1 centres in diamond on resonance with a probe NV spin. The first column indicates the $^{14}$N nuclear spin projection, $m_I^{(\rm t)}$, for the single-quantum transitions, and the initial and final projections for the double-quantum transitions. The second column indicates the symmetry axis of the P1 centre, along the $[111]$ axis (on-axis) or along one of the other three crystallographic axes (off-axis). The theoretical values are obtained as described in Sec. \ref{sec:methods:P1}, with uncertainties estimated based on the uncertainty on the value of $D_{\rm NV}$, which is the dominant source of error here. The experimental values are extracted from fitting the spectra in Figs. \ref{Fig2}a and \ref{Fig2}b with a sum of Lorentzian functions, with the uncertainty indicated being the standard error given by the fit.}	
\label{Table:EPR}	
\end{table}

\begin{figure}[b]
\begin{center}
\includegraphics[width=0.49\textwidth]{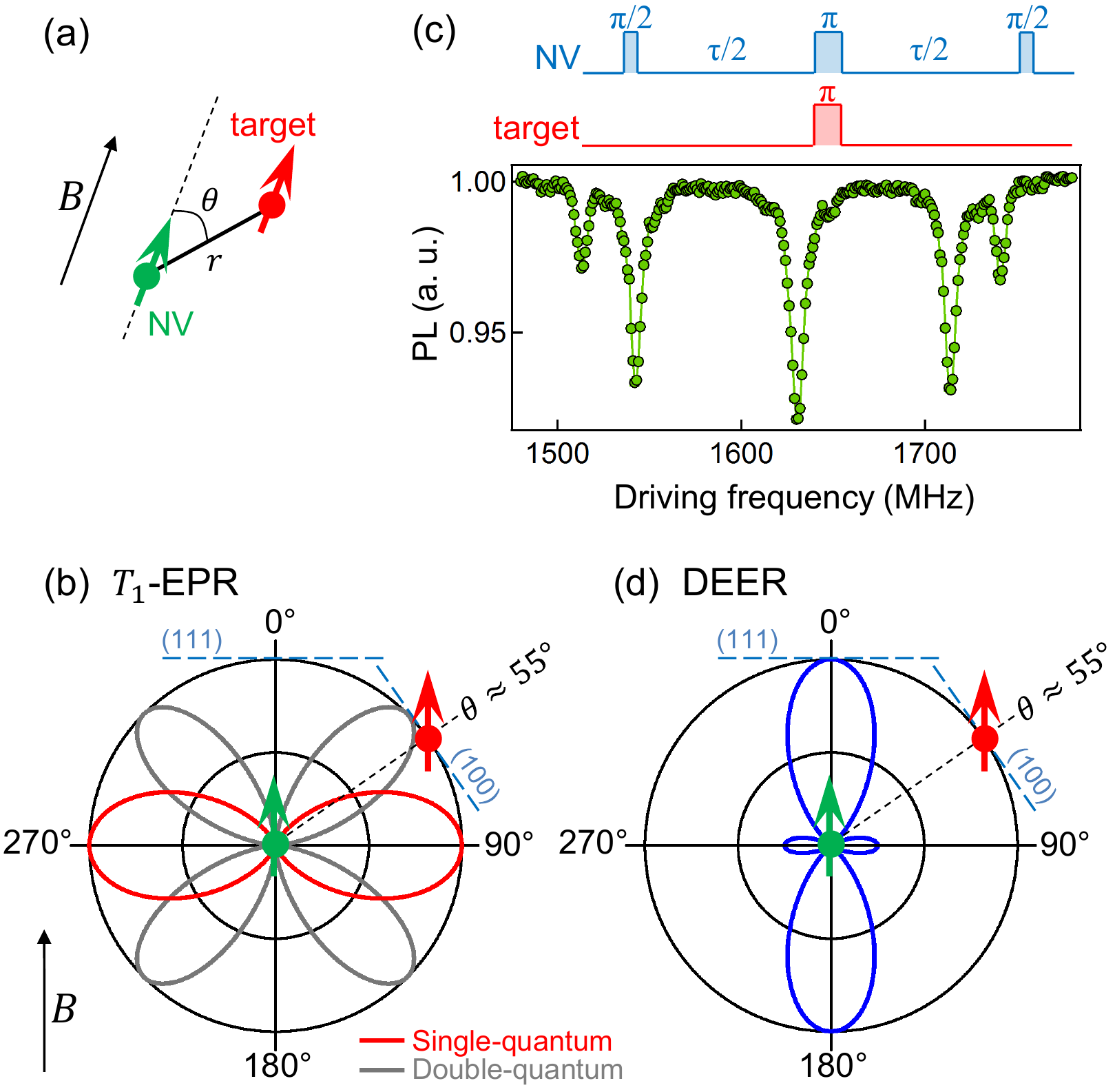}
\caption{(a) The target spin is located at a distance $r$ from the NV spin, the NV-target direction forming an angle $\theta$ with the external magnetic field. (b) Polar plot of the interaction strength of the $T_1$-EPR technique as a function of $\theta$ for single-quantum (red) and double-quantum (grey) transitions. For each case the strength is normalised by the maximum value, so that the outer circle corresponds to maximum strength. Blue dashed lines depict the (100) and (111) diamond surfaces for a target spin located on the surface right above the NV spin, assuming the NV to be oriented along the [111] crystallographic axis. This corresponds to angles $\theta\approx55^\circ$ and $\theta=0^\circ$, respectively. (c) Double electron-electron resonance (DEER) spectrum recorded at $B\approx580$ G using the same NV probe as in Fig. \ref{Fig2}. The sequence of microwave pulses is shown above the graph, in blue for the pulses on resonance with the NV spin, in red for the dark spins (here the P1 centres). The evolution time of the spin echo sequence is $\tau=1.5~\mu$s. (d) Polar plot of the signal intensity of the DEER technique as a function of the angle $\theta$, normalised by the maximum value. The signal vanishes at $\theta\approx55^\circ$, i.e. for a target spin on a (100) surface.} 
\label{Fig3}
\end{center}
\end{figure}

To gain further insight into the origin of the increased NV relaxation rate, we measured full $T_1$ relaxation curves for two different NV transition frequencies, $\omega_{\rm NV}=1300$ MHz where no resonance with the P1 centres is observed, and $\omega_{\rm NV}=1380$ MHz which corresponds to the single-quantum transition with $m_I^{(\rm t)}=-1$ of the on-axis P1 centres (Fig. \ref{Fig2}d). While the off-resonance data shows a single exponential behaviour with a characteristic decay rate $\Gamma_{1,\rm ph}=220(20)$ s$^{-1}$ associated with phonon-dominated relaxation \cite{Jarmola2012}, the on-resonance data reveals a bi-exponential behaviour. This is the signature of a resonance process, where only one of the NV transitions (here $m_S^{(\rm p)}=0\rightarrow -1$) is driven by the environment while the other transition ($m_S^{(\rm p)}=0\rightarrow +1$) remains unaffected. Assuming a simple three-level rate equation model in which the interaction with the target spins creates an additional near-resonance relaxation channel for one of the NV spin transitions with a transition rate $\Gamma_{1,\rm res}$, one finds that the relaxation curve takes the bi-exponential form
\begin{equation} \label{eq:T1curve}
I_s(\tau)=I_\infty\left[1+\frac{\cal C}{4}\left(e^{-\Gamma_{1,\rm ph}\tau}+3e^{-(\Gamma_{1,\rm ph}+\Gamma_{1,\rm res})\tau}\right)\right]
\end{equation}
where $I_\infty$ and ${\cal C}$ are constants (see derivation in Sec. \ref{sec:methods:T1curve}). Fitting this equation to the data of Fig. \ref{Fig2}d while fixing $\Gamma_{1,\rm ph}=220$ s$^{-1}$ yields $\Gamma_{1,\rm res}=6.2(6)\times10^3$ s$^{-1}$. 

The spin relaxation rate caused by a single target electron spin at resonance, located at a distance $r$ from the NV spin and forming an angle $\theta$ with the external magnetic field (Fig. \ref{Fig3}a), is given by
\begin{equation} 
\label{eq:GintP1}
\Gamma_{1,\rm res}^{\rm EPR} = \frac{1}{\Gamma_{2}}\left(\frac{\mu_0 \tilde{\gamma}_{\rm NV} \tilde{\gamma}_{e} h}{2 \sqrt{2}}\right)^2 \left(\frac{3\sin^2\theta}{r^3}\right)^2
\end{equation}
where $\Gamma_{2}$ is the total dephasing rate of the NV-target spin system (see derivation in Sec. \ref{sec:methods:GammaInt}). In the experiment, a given P1 centre is on resonance with the NV only a small fraction of the time for a given resonant magnetic field, due to the four equiprobable symmetry axes, two electron spin states and three nuclear spin states. This results in an effective relaxation rate weaker than that stated by Eq. (\ref{eq:GintP1}), e.g. by a factor 24 for the transition probed in Fig. \ref{Fig2}d. Taking this into account, and assuming a typical dephasing rate $\Gamma_{2}=10^6$ s$^{-1}$ and angle $\theta=\pi/4$, we infer a distance $r\approx10$ nm to the nearest neighbouring P1 centre, which contribute most to the signal \cite{Hall2015}. 

The double-quantum transitions occur via a two-step process, e.g. $|0,+1/2,0\rangle\rightarrow|0,-1/2,+1\rangle\rightarrow|-1,-1/2,+1\rangle$, where the ket refers to the full state $|m_S^{\rm (p)},m_S^{\rm (t)},m_I^{\rm (t)}\rangle$. The first step is enabled by transverse hyperfine coupling within the P1 centre (strength $A_\perp$), while the second step is enabled by dipolar interaction between the NV and P1 electron spins (strength $\omega_d$). However, although the initial and final state have the same energy at the resonant field $B_{\rm res}$, the intermediate state is detuned by an energy dominated by the electron Zeeman shift $\omega_e=-\tilde{\gamma}_eB_{\rm res}$. Consequently, the NV relaxation rate is expected to scale as $(A_\perp\omega_d/\omega_e)^2$. Precisely, from a full analysis of the NV-P1 interaction (see Sec. \ref{sec:methods:P1strength}) we find an on-resonant relaxation rate
\begin{eqnarray} 
\label{eq:GintP1dbl}
\Gamma_{1,\rm res}^{\rm EPR,double} &\approx& \frac{1}{\Gamma_{2}}\left(\frac{A_\perp}{\omega_e}\right)^2 \left(\frac{\mu_0 \tilde{\gamma}_{\rm NV} \tilde{\gamma}_{e} h}{2\sqrt{2}}\right)^2 \left(\frac{3\sin2\theta}{r^3}\right)^2 \nonumber \\
&\approx& \left(\frac{A_\perp}{\omega_e}\right)^2 \left(\frac{\sin2\theta}{\sin^2 \theta}\right)^2 \Gamma_{1,\rm res}^{\rm EPR}.
\end{eqnarray}
At the field where these resonances occur ($B\approx500$ G), the prefactor is $(A_\perp/\omega_e)^2\approx3\times10^{-3}$, which is why the double-quantum transitions appear significantly weaker than the single-quantum transitions in the $T_1$-EPR spectrum. Apart from this suppression factor, the angular dependence also differs from that of the single-quantum transitions, because they rely on a different term of the dipolar interaction. This is illustrated in Fig \ref{Fig3}b. For the situation where the relaxation is dominated by a single target electron spin, this provides a way to extract the position $\left(r,\theta\right)$ of the target relative to the probe. This opens the possibility of extracting spatial information on a target surface electron-nuclear spin system with a hyperfine splitting such as nitroxide spin labelled proteins \cite{Shi2015}. In the present experiment, it was not feasible to implement such a spatial resolution scheme as several P1 centres contribute to the signal. 

In order to contrast $T_1$-EPR with the most common method of $T_2$-EPR, a DEER spectrum was obtained from the same NV probe. This is shown in Fig. \ref{Fig3}c which exhibits only the five transitions associated with the single-quantum P1 transitions as seen in the $T_1$-EPR spectrum. The double-quantum transitions within the P1 centre are not addressable via a simple oscillating magnetic field and are therefore not seen in the DEER spectrum \cite{Hall2015}. In addition to probing transitions that DEER is unable to, $T_1$-EPR could also act in a complementary fashion to DEER due to their different angular dependences. As shown in Eq. (\ref{eq:GintP1}), $T_1$-EPR is sensitive to interactions with an angular dependence of $\sin^4\theta$ (single-quantum transitions). In contrast, DEER detects a different term of the dipole-dipole interaction between the NV probe and the target spin, which gives an angular dependence of $\left[1-3\cos^2\left(\theta \right)\right]^2$ as illustrated in Fig \ref{Fig3}d. The most common surface orientation for single crystal diamond samples is (100), where DEER has no sensitivity to spins on the surface directly above the NV, whereas $T_1$-EPR has non-zero sensitivity for such spins. On the other hand, for (111) surfaces, $T_1$-EPR has no sensitivity to spins located above the NV while DEER is at a maximum. Each technique is sensitive to spins at different angles and hence could be used in a complementary fashion to map the entire region around a single NV probe.

\subsection{$T_1$-NMR spectroscopy of P1 centres}

We now turn to the NMR transitions of the P1 centres, in which only the nuclear spin projection ($|\Delta m_I^{(\rm t)}|=1$) changes, whilst the electron spin projection is conserved ($\Delta m_S^{(\rm t)}=0$). These occur at transition frequencies from 40 to 60 MHz on either side of the GSLAC of the NV centre, determined by the hyperfine, quadrupole, and Zeeman couplings of the P1 nuclear spins. Assuming direct dipole-dipole interaction between the NV spin and the P1 nuclear spin, the NV relaxation rate at resonance can be predicted using Eq. (\ref{eq:GintP1}) by simply replacing the electron gyromagnetic ratio, $\tilde{\gamma}_e$, by the $^{14}$N gyromagnetic ratio, $\tilde{\gamma}_N$. This leads to a decay rate weaker by a factor of $(\tilde{\gamma}_N/\tilde{\gamma}_e)^2\approx10^{-8}$ than for the EPR transitions. 

However, in the present case there exists another interaction mechanism that leads to a greatly enhanced relaxation rate, which we will refer to as hyperfine-enhanced NMR. Similar to double-quantum EPR, hyperfine-enhanced NMR  occurs via a two-step process, here involving a double flip of the P1 electron spin, e.g. $|0,+1/2,0\rangle\rightarrow|0,-1/2,+1\rangle\rightarrow|-1,+1/2,+1\rangle$, using the notation $|m_S^{\rm (p)},m_S^{\rm (t)},m_I^{\rm (t)}\rangle$. In this example, the first step is enabled by transverse hyperfine coupling within the P1 centre (strength $A_\perp$), while the second step is enabled by dipolar coupling between the NV and P1 electron spins. The intermediate state is detuned from the initial and final states by an energy dominated by the electron Zeeman shift $\omega_e=-\tilde{\gamma}_eB_{\rm res}$. Full analysis of the NV-P1 interaction (see Sec. \ref{sec:methods:P1strength}) gives the on-resonant relaxation rate
\begin{eqnarray} 
\label{eq:GintP1NMR}
\Gamma_{1,\rm res\pm}^{\rm NMR,hyp} &\approx& \frac{1}{\Gamma_{2}}\left(\frac{A_\perp}{\omega_e}\right)^2 \left(\frac{\mu_0 \tilde{\gamma}_{\rm NV} \tilde{\gamma}_{e} h}{2\sqrt{2}}\right)^2 \nonumber \\
& & \times \left(\frac{3 \sin^2\theta-1\pm1}{r^3}\right)^2 \\
&\approx& \left(\frac{A_\perp}{\omega_e}\right)^2 \left(\frac{3 \sin^2\theta-1\pm1}{3 \sin^2\theta}\right)^2 \Gamma_{1,\rm res}^{\rm EPR} \nonumber
\end{eqnarray}
where the $\pm$ sign depends whether the dipole-dipole transition goes from $m_S^{\rm (t)}=+1/2$ to $-1/2$ (`+' sign) or from $-1/2$ to $+1/2$ (`-' sign). At the field where the transitions occur ($B\approx1000$ G), the suppression factor in Eq. (\ref{eq:GintP1NMR}) is $(A_\perp/\omega_e)^2\approx10^{-3}$. As a result, the resonances in hyperfine-enhanced NMR are expected to be $\approx10^{-3}$ weaker than in single-quantum EPR, but only 4 times weaker than in double-quantum EPR (due to the field being twice as large) and, crucially, $\approx10^{5}$ stronger than if probed via direct dipolar interaction between the NV electron spin and the P1 nuclear spin. In semi-classical terms, hyperfine-enhanced NMR can be interpreted as a modulation of the P1 electron spin precession caused by the hyperfine-coupled P1 nuclear spin precession, which acts as a beat frequency in the NV-P1 electron-electron interaction.

\begin{figure*}[t]
\begin{center}
\includegraphics[width=0.94\textwidth]{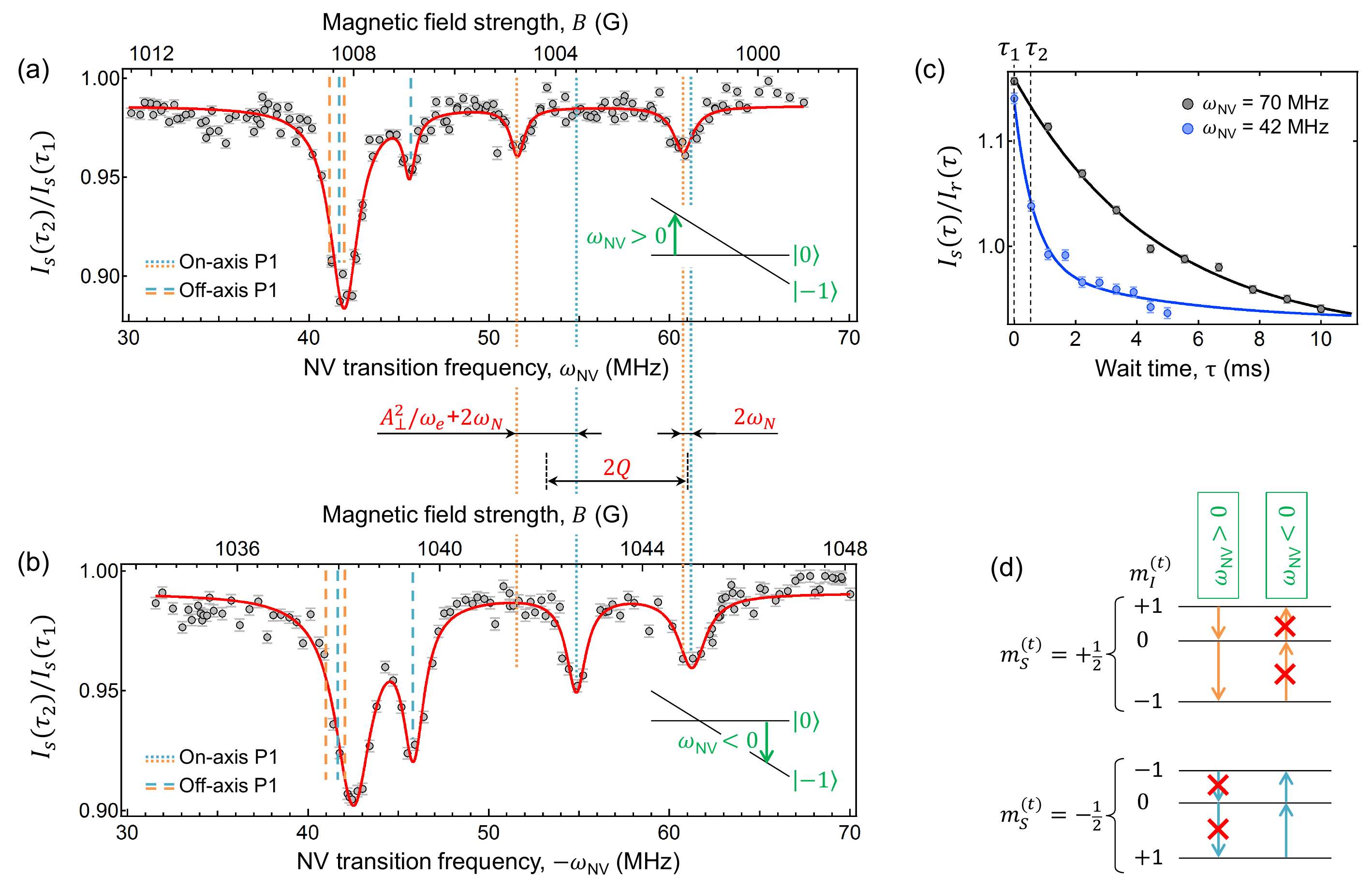}
\caption{(a,b) $T_1$-NMR spectra of P1 centres in diamond obtained by measuring the population decay of a single NV spin after a wait time $\tau_2=600~\mu$s. The normalised PL signal $I_s(\tau_2)/I_s(\tau_1)$ with $\tau_1=1~\mu$s is plotted against the NV transition frequency $\omega_{\rm NV}$ obtained from the ODMR spectrum. In (a) $\omega_{\rm NV}>0$, i.e., the magnetic field is $B<1024$ G (before GSLAC, see inset), while in (b) $\omega_{\rm NV}<0$ and $B>1024$ G (see inset). In (a,b), solid red curves are data fitting to a sum of four Lorentzian functions, and vertical lines indicate the theoretical frequencies with colours as defined in (d). (c) Full $T_1$ relaxation curve measured off resonance (black markers, $\omega_{\rm NV}=70$ MHz) and on a particular resonance (blue markers, $\omega_{\rm NV}=42$ MHz). Solid lines are data fitting to Eq. (\ref{eq:T1curve}). The vertical dashed line indicates the probe times $\tau_1$ and $\tau_2$ used in (a,b). (d) Energy levels of a single P1 centre showing the NMR transitions ($\Delta m_S=0$, $|\Delta m_I|=1$). Orange and blue arrows correspond to transitions within the $m_S=+1/2$ and $m_S=-1/2$ manifolds, respectively. Downward (upward) arrows correspond to $\omega_{\rm NV}>0$ ($\omega_{\rm NV}<0$). Red crosses indicate transitions with the weakest relaxation rate, $\Gamma_{1,\rm res-}^{\rm NMR,hyp}$.} 
\label{Fig4}
\end{center}
\end{figure*}

To probe the NMR transitions experimentally, we chose a longer probe evolution time, $\tau_2 = 600~\mu$s and normalised the signal using a reference probe evolution time $\tau_1 = 1~\mu$s. The $T_1$-NMR spectra recorded in the range 30-70 MHz before the GSLAC $\left(\omega_{\rm NV} > 0\right)$ and after the GSLAC $\left(\omega_{\rm NV} < 0\right)$ are shown in Figs. \ref{Fig4}a and \ref{Fig4}b, respectively, each revealing four well-resolved peaks. A full $T_1$ relaxation curve recorded at one of the resonances ($\omega_{\rm NV} = +42$ MHz, Fig. \ref{Fig4}c) confirms that the interaction is significantly weaker than in the $T_1$-EPR spectrum, with an induced decay rate $\Gamma_{1,\rm res} = 1.4(2) \times 10^3$~s$^{-1}$, but is far greater than if probed via direct dipole-dipole coupling alone.

\begin{table}[t]
(a) Before GSLAC ($\omega_{\rm NV}>0$)
\begin{tabular}{|c|c|c|c|c|}
\hline
Symmetry & \multirow{2}{*}{$m_S^{(\rm t)}$} & \multirow{2}{*}{$m_I^{(\rm t)}$} & \multicolumn{2}{c|}{$\omega_{\rm NV}(B_{\rm res})$ (MHz)} \\
\cline{4-5}
(on/off axis) & & & Theory & Experiment \\
\hline 
\hline
\multirow{4}{*}{on} & \multirow{2}{*}{$+1/2$} & $+1\rightarrow0$ & 51.53(1) & 51.6(1) \\
\cline{3-5}
&  & $0\rightarrow-1$ & 60.64(1) & \multirow{2}{*}{60.8(1)} \\
\cline{2-4}
& \multirow{2}{*}{$-1/2$} & $0\rightarrow+1$ & 61.26(1)  &  \\
\cline{3-5}
&  & $-1\rightarrow0$ & 54.47(1) & NR \\
\hline
\multirow{4}{*}{off} & \multirow{2}{*}{$+1/2$} & $+1\rightarrow0$ & 41.86(1) & \multirow{3}{*}{42.0(1)} \\
\cline{3-4}
&  & $0\rightarrow-1$ & 40.85(1) &  \\
\cline{2-4}
& \multirow{2}{*}{$-1/2$} & $0\rightarrow+1$ & 41.47(1)  &  \\
\cline{3-5}
&  & $-1\rightarrow0$ & 45.73(1) & 45.6(1) \\
\hline
\end{tabular}

\hspace{10cm}

(b) After GSLAC ($\omega_{\rm NV}<0$)
\begin{tabular}{|c|c|c|c|c|}
\hline
Symmetry & \multirow{2}{*}{$m_S^{(\rm t)}$} & \multirow{2}{*}{$m_I^{(\rm t)}$} & \multicolumn{2}{c|}{$\omega_{\rm NV}(B_{\rm res})$ (MHz)} \\
\cline{4-5}
(on/off axis) & & & Theory & Experiment \\
\hline 
\hline
\multirow{4}{*}{on} & \multirow{2}{*}{$+1/2$} & $0\rightarrow+1$ & 51.59(1) & NR \\
\cline{3-5}
&  & $-1\rightarrow0$ & 60.65(1) & \multirow{2}{*}{61.2(1)} \\
\cline{2-4}
& \multirow{2}{*}{$-1/2$} & $+1\rightarrow0$ & 61.29(1)  &  \\
\cline{3-5}
&  & $0\rightarrow-1$ & 54.48(1) & 54.9(1) \\
\hline
\multirow{4}{*}{off} & \multirow{2}{*}{$+1/2$} & $0\rightarrow+1$ & 41.92(1) & \multirow{3}{*}{42.5(1)} \\
\cline{3-4}
&  & $-1\rightarrow0$ & 40.85(1) &  \\
\cline{2-4}
& \multirow{2}{*}{$-1/2$} & $+1\rightarrow0$ & 41.49(1)  &  \\
\cline{3-5}
&  & $0\rightarrow-1$ & 45.70(1) & 45.8(1) \\
\hline
\end{tabular}
\caption{Summary of the analytic and experimental NMR transition frequencies of P1 centres in diamond on resonance with a probe NV spin before (a) and after (b) the GSLAC. The analytic values are obtained from Eq. (\ref{eq:rescond}) using the Hamiltonian (\ref{eq:HP1}) (see Sec. \ref{sec:methods:P1}), with uncertainties estimated based on the uncertainties on the hyperfine parameters \cite{Cook1966}. The experimental values are extracted from the spectra in Figs. \ref{Fig3}a and \ref{Fig3}b. The first column indicates the symmetry axis of the P1 centre, along the $[111]$ axis (on-axis) or along one of the other three crystallographic axes (off-axis). The second column indicates the electron spin projection ($m_S^{(\rm t)}$) while the third column indicates the transition for the $^{14}$N nuclear spin projection ($m_I^{(\rm t)}$).  NR, not resolved.}	
\label{Table:NMR}	
\end{table} 

Each family of P1 centres (on and off axis) gives rise to four nuclear transitions on either side of the GSLAC (Fig. \ref{Fig3}d), resulting in a maximum of eight transitions in both spectra. The theoretical values obtained from the Hamiltonian (Eq. (\ref{eq:HP1})) are given in Table \ref{Table:NMR} and indicated as vertical lines in Figs. \ref{Fig3}a and \ref{Fig3}b, showing excellent agreement with the values obtained from the experimental spectra. However, not all nuclear transitions are resolved experimentally, which is clearer for the on-axis P1 centres. This is due to the two possible signs in Eq. (\ref{eq:GintP1NMR}), which results in different relaxation rates depending on the transition considered. Averaging over the position of the P1 centre (via the angle $\theta$), it can be shown that $\langle\Gamma_{1,\rm res+}^{\rm NMR,hyp}\rangle= 6\langle\Gamma_{1,\rm res-}^{\rm NMR,hyp}\rangle$ (see Sec. \ref{sec:NMRsuppressed}). The transitions with the smallest decay rate, $\Gamma_{1,\rm res-}^{\rm NMR,hyp}$, are indicated in \ref{Fig4}d (red crosses), and match the transitions that are not seen experimentally with the probe time used, $\tau_2=600~\mu$s. A longer probe time could be employed to detect those weaker transitions, however the sensitivity decreases as $\tau$ approaches the background relaxation time $T_{1,\rm ph}\approx 5$ ms.    

For the on-axis P1 centres,  the four resonance frequencies within the P1 system are
\begin{align} \label{eq:NMRtansitions}
\begin{split}
\omega_{\rm t}\left({\left\vert+\tfrac{1}{2},0\right\rangle\leftrightarrow\left\vert+\tfrac{1}{2},+1\right\rangle}\right) &=  \frac{A_\parallel}{2} -\frac{A_\perp^2}{2\omega_e}+Q-\omega_N \\
\omega_{\rm t}\left({\left\vert+\tfrac{1}{2},0\right\rangle\leftrightarrow\left\vert+\tfrac{1}{2},-1\right\rangle}\right) &= \frac{A_\parallel}{2} -Q-\omega_N \\
\omega_{\rm t}\left({\left\vert-\tfrac{1}{2},0\right\rangle\leftrightarrow\left\vert-\tfrac{1}{2},+1\right\rangle}\right) &= \frac{A_\parallel}{2} -Q+\omega_N \\
\omega_{\rm t}\left({\left\vert-\tfrac{1}{2},0\right\rangle\leftrightarrow\left\vert-\tfrac{1}{2},-1\right\rangle}\right) &= \frac{A_\parallel}{2} +\frac{A_\perp^2}{2\omega_e}+Q+\omega_N \\
\end{split}
\end{align}
where $\omega_e=-\tilde{\gamma}_eB_{\rm res}$ and $\omega_N=-\tilde{\gamma}_NB_{\rm res}$ are the electron and nuclear Zeeman shifts at the corresponding resonant field, and the kets denote the P1 states $\vert m_S^{(\rm t)},m_I^{(\rm t)}\rangle$. The double arrow $\leftrightarrow$ accounts for the transitions both before and after the GSLAC, leading to eight resonances in total. From the experimental spectra, all frequency values of the four strongest transitions (with decay rate $\Gamma_{1,\rm res+}^{\rm NMR,hyp}$) can be determined. Therefore, one can use Eqs. (\ref{eq:NMRtansitions}) to directly deduce the values of $A_\parallel$, $A_\perp$, $Q$ and $\tilde{\gamma}_N$, as illustrated in Figs. \ref{Fig3}a and \ref{Fig3}b. Here we find $A_\parallel=114.2(1)$ MHz, $A_\perp=91(3)$ MHz, $Q=-3.9(1)$ MHz and $\tilde{\gamma}_N=1.9(5)$ MHz/T. These values are all in good agreement with the literature values obtained from ensemble measurements \cite{Cook1966}. This illustrates that our technique has the capacity to measure the hyperfine and quadrupole coupling parameters as well as the gyromagnetic ratio of the target system, at the single atom level.

\section{Towards $T_1$-NMR at the single nuclear spin level} \label{sec:NMR}

\begin{figure*}[t]
\begin{center}
\includegraphics[width=0.94\textwidth]{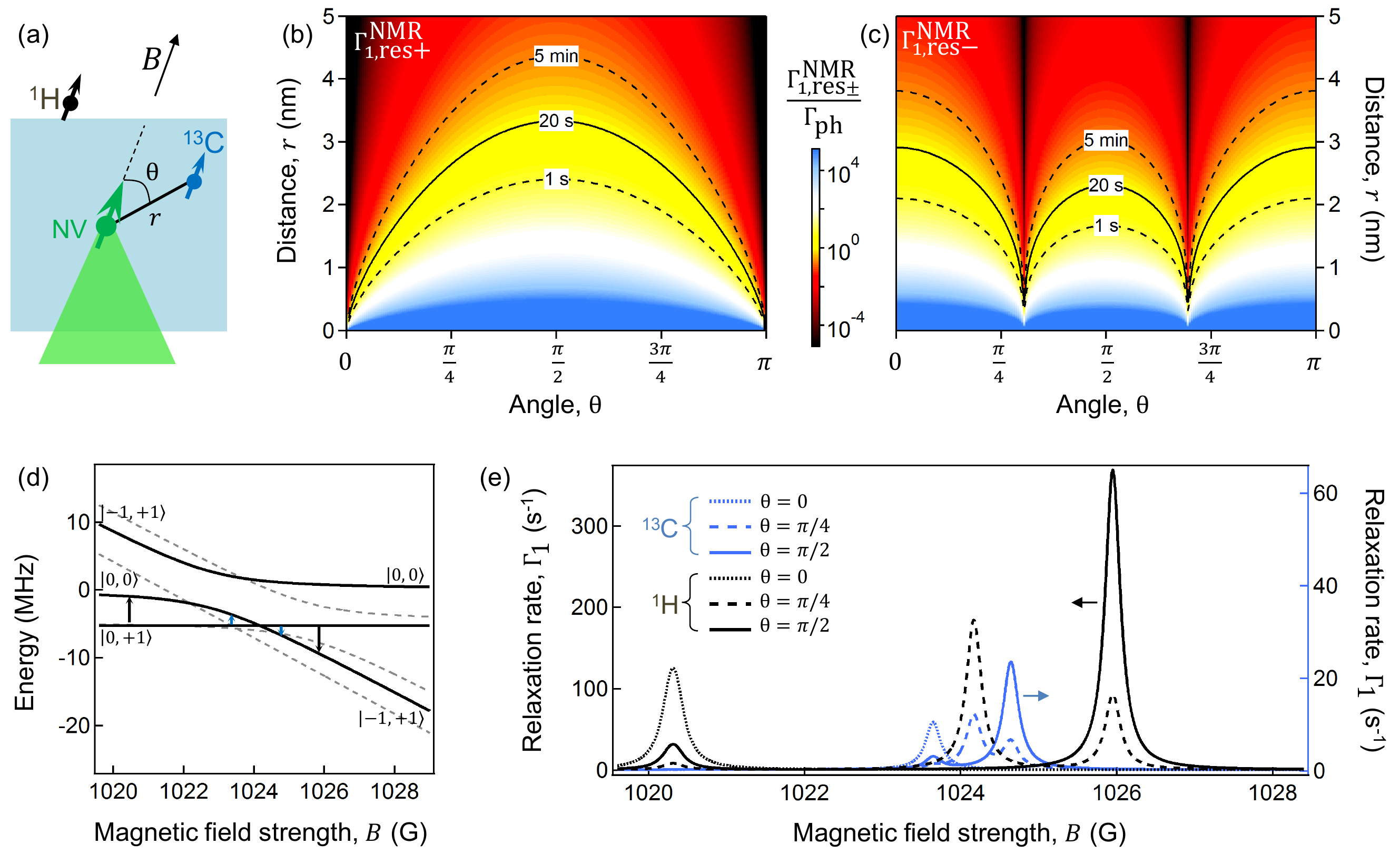}
\caption{(a) The target spins considered are single $^1$H and $^{13}$C nuclear spins, either inside or outside the diamond lattice. $\theta$ refers to the polar angle and $r$ to the distance from the NV probe to the relevant nuclear spin. (b,c) Relaxation rate $\Gamma_{1,\rm res\pm}^{\rm NMR}$ induced by resonant interaction with a single $^1$H nuclear spin, calculated as a function of $r$ and $\theta$ at the resonant field $B_{\rm res}^+$ in (b) and $B_{\rm res}^-$ in (c). The values are normalised by the background relaxation rate set to be $\Gamma_{1,\rm ph}=200$ s$^{-1}$ and the ratio $\Gamma_{1,\rm res\pm}^{\rm NMR}/\Gamma_{1,\rm ph}$ is plotted in log scale. Black lines are contours corresponding to $\Gamma_{1,\rm res\pm}^{\rm NMR}/\Gamma_{1,\rm ph}=0.2$, 1 and 7 from top to bottom. This translates into a total acquisition time of 5 min, 20 s and 1 s, respectively, in order to detect the interaction with a signal-to-noise ratio of 1. (d) Energy levels of the NV spin including the hyperfine sublevels, as a function of $B$. Solid lines correspond to the levels that are populated in the experiment. The arrows indicate the transitions on resonance with a $^1$H spin (black) and a $^{13}$C spin (blue). (e) Simulated $T_1$-NMR spectrum of a single $^1$H spin (red lines) or a single $^{13}$C spin (blue lines). The relaxation rate $\Gamma_{1,\rm res}$ is calculated as a function of the magnetic field strength $B$. The target spin is assumed to be at a distance $r=3$ nm from the NV spin.} 
\label{Fig5}
\end{center}
\end{figure*}

In the previous section, we reported successful experimental detection of the EPR and NMR transitions of a small ensemble of P1 centres surrounding a single NV spin probe. Here the strength of the NMR transitions was enhanced by the hyperfine interaction within the target system. Based on these results, we now analyse theoretically the possibility of performing $T_1$-NMR spectroscopy on nuclear spin systems free of electron spins, that is, a target with no significant hyperfine interaction to an environmental electron spin other than the probe. We first consider a single nuclear spin-1/2 with gyromagnetic ratio $\tilde{\gamma_{\rm t}}$ located a distance $r$ from the NV spin and forming an angle $\theta$ with the direction of the external magnetic field (see Fig. \ref{Fig5}a). The eigenstates of the non-interacting system are labeled as $\vert m_S^{(\rm p)},m_I^{(\rm t)}\rangle$ where $m_S^{(\rm p)}$ and $m_I^{(\rm t)}$ are the projection of the NV electron spin probe and target nuclear spin, respectively. Since $|\tilde{\gamma_{\rm t}}|<|\tilde{\gamma_{\rm NV}}|$, there are two resonances at magnetic field strengths $B_{\rm res}^\pm$ given by Eq. (\ref{eq:Bres}) where $D_{\rm t}=0$. In this basis, the first resonance ($B_{\rm res}^-$) corresponds to the transition $|0,-1/2\rangle\rightarrow|-1,+1/2\rangle$ while the second one ($B_{\rm res}^+$) corresponds to $|0,+1/2\rangle\rightarrow|-1,-1/2\rangle$, assuming $\tilde{\gamma_{\rm t}}>0$. By solving the evolution of the system starting in either the state $|0,-1/2\rangle$ or $|0,+1/2\rangle$, one can express the relaxation rate $\Gamma_{1,\rm res\pm}^{\rm NMR}$ induced on the NV spin when each resonance condition is fulfilled. We obtain (see Sec. \ref{sec:methods:GammaInt})
\begin{align} \label{eq:GammaInt+-}
\Gamma_{1,\rm res\pm}^{\rm NMR} &= \frac{1}{\Gamma_{2}}\left(\frac{\mu_0\tilde{\gamma}_{\rm NV}\tilde{\gamma}_{\rm t} h}{2\sqrt{2}} \right)^2 \left(\frac{3\sin^2\theta-1\pm1}{r^3} \right)^2
\end{align}
where $\Gamma_{2}$ is the total dephasing rate of the system. 

As a prototypical system, we consider a proton ($^1$H) spin, which has a gyromagnetic ratio $\tilde{\gamma}_{\rm t}=42.58$ MHz/T. The two resonances with the NV spin occur at fields $B_{\rm res}^\pm-B_{\rm GSLAC}\approx\pm2$ G from the GSLAC. The dephasing rate is assumed to be dominated by that of the NV spin since nuclear spins interact much more weakly with their environment than electron spins. We will take $\Gamma_{2}=10^6$ s$^{-1}$, which corresponds to typical dephasing rate for a near-surface NV centre \cite{Ohashi2013}. Figs. \ref{Fig4}b and \ref{Fig4}c show the normalised relaxation rates $\Gamma_{1,\rm res\pm}^{\rm NMR}/\Gamma_{1,\rm ph}$ computed as a function of $r$ and $\theta$ using the above parameters along with a background phonon relaxation rate for the NV of $\Gamma_{1,\rm ph}=200$ s$^{-1}$. The solid black line indicates the contour $\Gamma_{1,\rm res\pm}^{\rm NMR}/\Gamma_{1,\rm ph}=1$, showing that the induced relaxation rate $\Gamma_{1,\rm res\pm}^{\rm NMR}$ reaches the phonon background relaxation rate $\Gamma_{1,\rm ph}$ for distances as large as 3 nm at $\theta=\pi/2$ for $\Gamma_{1,\rm res+}^{\rm NMR}$ and at $\theta=0,\pi$ for $\Gamma_{1,\rm res-}^{\rm NMR}$. 

To estimate the acquisition time that would be required experimentally to detect a single proton spin, we need to compare the change of PL signal at resonance with the measurement noise. Assuming the noise is dominated by photon shot noise, the signal-to-noise ratio when measuring the PL signal after a wait time $\tau$ (i.e., $I_s(\tau)$) can be expressed as (see Sec. \ref{sec:methods:sensitivity})
\begin{align} \label{eq:SNR}
{\rm SNR}(\tau)\approx\sqrt{\frac{{\cal R}t_{\rm ro}T_{\rm tot}}{\tau}}\frac{3{\cal C}}{4}e^{-\Gamma_{1,\rm ph}\tau}\left(1-e^{-\Gamma_{1,\rm res}\tau}\right)
\end{align}
where ${\cal R}$ is the photon count rate under continuous laser excitation, $t_{\rm ro}$ is the read-out time, $T_{\rm tot}$ is the total acquisition time and ${\cal C}$ is the $T_1$ contrast as defined in Eq. (\ref{eq:T1curve}). In the limit $\Gamma_{1,\rm res}\ll\Gamma_{1,\rm ph}$, the wait time that maximises ${\rm SNR}(\tau)$ is $\tau_{\rm opt}=(2\Gamma_{1,\rm ph})^{-1}$, however in general the optimum wait time $\tau_{\rm opt}$ is smaller and depends on $\Gamma_{1,\rm res}$. For a given induced relaxation rate $\Gamma_{1,\rm res}$, we define the minimum acquisition time $T_{\rm tot,min}$ as the time needed to obtain an optimised ratio ${\rm SNR}(\tau_{\rm opt})$ equal to unity. Using typical experimental conditions, namely ${\cal R}=2\times10^5$ s$^{-1}$, $t_{\rm ro}=300$ ns and ${\cal C}=0.25$, we find that 20 s are required to detect an interaction such that $\Gamma_{1,\rm res}=\Gamma_{1,\rm ph}$. In Figs. \ref{Fig4}b and \ref{Fig4}c two other contours are shown corresponding to an acquisition time of 1 s and 5 minutes. The latter case enables the detection of a proton spin located up to 4 nm away from the NV probe while still allowing acquisition of a full spectrum in a few hours.

In calculating the interaction strength above, the hyperfine interaction of the NV electron spin with its $^{14}$N (or $^{15}$N) nucleus was neglected. However, the resonances with a $^1$H spin occur at magnetic field strengths close enough to the GSLAC so that hyperfine-induced spin mixing may become significant and affect the resonant fields as well as the interaction strength. To account for this effect, we numerically simulated $T_1$-NMR spectra of a single $^1$H spin while considering the full hyperfine structure of the $^{14}$NV centre (see Sec. \ref{sec:methods:H1simu}). The energy levels of the NV centre near the GSLAC are shown in Fig. \ref{Fig4}d, where the non-perturbed eigenstates are labelled $|m_S^{(\rm p)},m_I^{(\rm p)}\rangle$ and $m_S^{(\rm p)}$ and $m_I^{(\rm p)}$ are the electron and nuclear spin projections of the NV, respectively. Under optical excitation for magnetic fields close to the GSLAC, the NV centre is efficiently prepared in the state $|0,+1\rangle$ \cite{Jacques2009,Fuchs2011}, which remains an eigenstate all across the GSLAC. Resonant interaction with the target spin can drive transitions from $|0,+1\rangle$ to $|-1,+1\rangle$. This state is mixed with the $|0,0\rangle$ state at the GSLAC owing to hyperfine coupling. In the simulation, the NV spin is initialised in the state $|0,+1\rangle$ while the target spin is initialised in a completely mixed state. The system's evolution is computed under the same assumptions as before, with a total dephasing rate $\Gamma_{2}=10^6$ s$^{-1}$. The decay of the population remaining in $|0,+1\rangle$ as the system evolves is then used to infer the relaxation rate $\Gamma_{1,\rm res}$. 

Fig. \ref{Fig4}e shows $\Gamma_{1,\rm res}$ as a function of the magnetic field strength $B$ across the GSLAC for a single $^1$H spin located at a distance $r=3$ nm with various angles $\theta$. Also shown for comparison is the case where the target is a single $^{13}$C spin, which is also a spin-1/2 but with a smaller gyromagnetic ratio. In both cases, three peaks are observed in the spectrum. The two side peaks correspond to the NV-target resonances and occur at fields $B_{\rm res}^+=1026.0$ G and $B_{\rm res}^-=1020.3$ G for the $^1$H case, and  $B_{\rm res}^+=1024.65$ G and $B_{\rm res}^-=1023.65$ G for the $^{13}$C case. The splitting between the two resonances is directly related to the gyromagnetic ratio according to Eq. (\ref{eq:Bres}) with a correction due to the level avoided crossing causing an asymmetry about the central feature. The comparison of the different angles $\theta$ illustrates the different angular dependences for the two resonances as expressed in Eq. (\ref{eq:GammaInt+-}). Thus, $T_1$-based NMR spectroscopy would enable not only identification of unknown spin species but also quantification of their densities, via the strength of decay and angular positions, by comparing the decay rate for each resonant transition. We note that the width of the resonances -- that is, the spectral resolution -- is given by the dephasing rate $\Gamma_{2}$ \cite{Hall2015}, in our case $\approx1$ MHz. It can therefore be improved by several orders of magnitude by engineering NV centres in high-purity diamond crystals \cite{Maurer2012}. 

The central feature common to the spectra of both species at $B=1024.17$ G is specific to the GSLAC of
the NV rather than the target spins themselves. It occurs at the magnetic field where the initial NV state $|0,+1\rangle$ crosses with one of the eigenstates containing a superposition of $|-1,+1\rangle$ and $|0,0\rangle$. Due to the $|0,+1\rangle\rightarrow|-1,+1\rangle$ transition being addressable via a resonant magnetic field, when degenerate, any non-axial static field will drive this transition. At this crossing, the non-axial component of the effective field produced by the target spin will drive this transition. In reality this crossing will be sensitive to any non-axial field caused by environmental noise rather than the target nuclear spins alone. Therefore this transition is not relevant for determining the nuclear spin species. 

\section{Conclusions} \label{sec:Conclusion}

In this work we have demonstrated a broadband, nanoscale method for interrogating environmental resonances of both nuclear and electronic species based on $T_1$-MR. Using a single spin probe in diamond we interrogated both the electronic and nuclear spin transitions of substitutional $^{14}$N impurities within the diamond lattice, showing this method's ability to sense in both the GHz and MHz regimes. Notably, we uncovered a hyperfine-enhanced mechanism for detecting nuclear transitions, which enhances the signal strength by several orders of magnitude over direct detection. The all-optical nature of the $T_1$-MR technique, freeing one from the requirement for microwave manipulation of either the probe or target, removes some issues associated with MR measurement techniques based on $T_2$ dephasing, as well as allowing the interrogation time to be extended from $T_2$ to $T_1$. Finally, we showed theoretically that this technique has the sensitivity to detect single proton spins at a distance of a few nm. $T_1$-MR thus provides a promising new avenue towards nanoscale nuclear magnetic resonance spectroscopy and imaging, across EPR and NMR regimes.

\section{Methods} \label{sec:methods}

\subsection{Experimental details}

\subsubsection{Experimental setup} \label{sec:methods:setup}

The experimental apparatus consists of a custom-built confocal microscope and a permanent magnet mounted on a scanning stage (Fig. \ref{Fig1}). The excitation source is a solid-state laser emitting at a wavelength $\lambda=532$ nm (Laser Quantum Gem 532). The objective lens (Olympus UPlanSApo 100x, NA = 1.4 Oil) is mounted on an XYZ scanning stage (PI P-611.3 NanoCube) to allow fast laser scanning. The PL emitted by the diamond sample is separated from the laser light using a dichroic beam splitter and a band-pass filter, and coupled into a multi-mode fibre connected to a single photon counting module (Excelitas SPCM-AQRH-14-FC). For $T_1$ measurements, the laser beam is modulated by an acousto-optic modulator (AA Opto-Electronic MQ180-A0,25-VIS) in a double pass configuration, and the PL signal is analysed by a time digitizer (FastComTec P7889). For ODMR measurements, a 20-$\mu$m copper wire is spanned on the surface of the diamond and connected to the output of a microwave generator (Agilent N5181A) modulated by a switch (Mini-Circuits ZASWA-2-50DR+). Laser and microwave modulations are controlled by a programmable pulse generator (SpinCore PulseBlasterESR-PRO 500 MHz). 

The sample is a type-Ib single crystal diamond grown by the High Pressure High Temperature (HPHT) synthesis process, purchased from Element Six, which has \{100\} oriented faces. The approximately 1 inch diameter cylindrical magnet is mounted such that its principal axis of magnetization is approximately parallel to the $[111]$ crystallographic axis of the diamond, which we refer to as the $z$ axis and corresponds to the symmetry axis of the investigated NV centres. The magnet is mounted on an XYZ scanning stage made of three linear translation stages (PI M-511). This allows movement of the magnet along the $z$ direction and thus variation of the strength $B$ of the magnetic field. For a given $z$ position, the field direction is finely aligned along the $z$ axis by moving the magnet in the $xy$ plane and maximising the PL intensity from the NV centre, as described in the next section. 

\subsubsection{Acquisition procedure} \label{sec:methods:procedure}

The spectra shown in Figs. \ref{Fig2} and \ref{Fig4} are obtained as follows. The magnet is stepped along the $z$ direction to vary the magnetic field strength $B$. For each magnet $z$ position, three operations are run consecutively, as described below.

\begin{figure}[b]
\begin{center}
\includegraphics[width=0.48\textwidth]{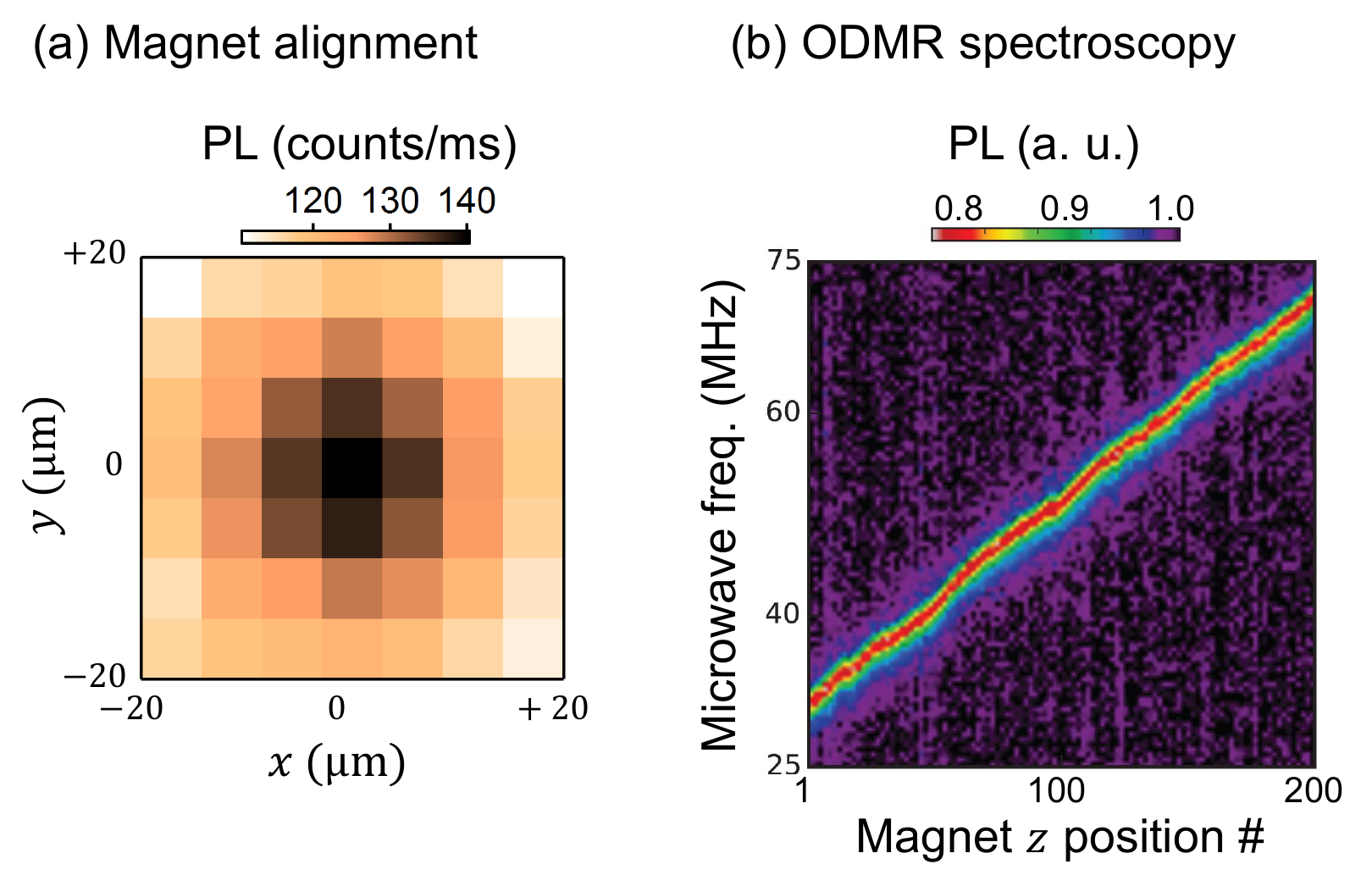}
\caption{(a) PL intensity of a single NV centre under continuous laser excitation as a function of the transverse position ($x,y$) of the magnet, at a magnetic field strength $B\approx1000$ G. The signal is maximum at the centre of the image, which corresponds to the case where the magnetic field is aligned with the NV symmetry axis. (b) Map of ODMR spectra recorded while stepping the magnet along the $z$ direction. The NV transition frequency varies approximately linearly with the $z$ position, here in the range $\omega_{\rm NV}=30-70$ MHz corresponding to $B=1000-1012$ G. } 
\label{Fig6}
\end{center}
\end{figure}

First, the magnet is scanned in the transverse $xy$ plane in order to fine tune the alignment of the field based on the PL intensity. Indeed, for field strengths above $\sim200$ G, the PL intensity quickly drops when the field direction is misaligned away from the NV centre's symmetry axis owing to spin mixing in the ground electronic state as well as in the excited state \cite{Epstein2005,Tetienne2012}. This effect is particularly pronounced around the GSLAC ($B\approx1024$ G) and around the excited state level anti-crossing (ESLAC, $B\approx512$ G), with a PL decrease of $\approx40\%$ with a misalignment angle of $>2^\circ$. Therefore, maximising the PL intensity gives us a way to precisely align the magnetic field along the $[111]$ direction. Fig. \ref{Fig6}a shows an example of PL intensity measured against the transverse position ($x,y$) of the magnet for a given $z$ position (here at a field strength of $B\approx1000$ G). We then set the transverse position of the magnet to the centre of a two-dimensional Gaussian function fitted to the data. Based on the uncertainty of the fit, we estimate the field to be aligned within $\pm1^\circ$ of the $[111]$ direction for the range of field strengths considered in this work.        

Second, an ODMR spectrum of the NV centre is recorded in order to determine the NV transition frequency $\omega_{\rm NV}$ and infer the field strength $B$. To this end, the PL intensity is measured while sweeping the microwave frequency across the $|0\rangle\rightarrow|-1\rangle$ resonance. To avoid power broadening \cite{Dreau2011}, 300-ns laser pulses are interleaved with 1-$\mu$s microwave pulses corresponding to a $\pi$-flip of the NV spin on resonance. The set of ODMR spectra recorded for the data of Fig. \ref{Fig3}a is shown in Fig. \ref{Fig6}b. Only one hyperfine transition of the NV centre is observed because of efficient polarisation of the $m_I^{(\rm p)}=+1$ nuclear spin state of the $^{14}$N nucleus intrinsic to the NV centre around the GSLAC and ESLAC \cite{Jacques2009,Fuchs2011}. The spectrum is fitted with a Lorentzian function to obtain the NV transition frequency $\omega_{\rm NV}$. The field strength $B$ is deduced using the relation $\omega_{\rm NV}=D_{\rm NV}-a_\parallel-\tilde{\gamma}_{\rm NV}B$ where $a_\parallel=-2.16$ MHz is the hyperfine coupling parameter of the NV centre (see further details in Sec. \ref{sec:methods:NV}).

Third, we apply a sequence of 3-$\mu$s laser pulses separated by different wait times $\tau$. The sequence is repeated typically $10^6$ times while the time-resolved PL is integrated. The resulting PL trace is then analysed to extract the quantities $I_s(\tau)$ and $I_r(\tau)$, which corresponds to the number of photons detected within a window of 300 ns at the start of the pulse and at the end of the pulse, respectively (see inset in Fig. \ref{Fig1}e). For the $T_1$-NMR spectra (Figs. \ref{Fig4}a and \ref{Fig4}b), there is no population decay at $\tau=1~\mu$s even on resonance, which allows us to normalise the signal as $I_s(\tau_2)/I_s(\tau_1)$ with a probe time $\tau_2=600~\mu$s and a normalisation time $\tau_1=1~\mu$s. For the $T_1$-EPR spectra however (Figs. \ref{Fig2}a and \ref{Fig2}b), the decay rate on resonance ($\Gamma_{1,\rm res}^{\rm EPR}$) is so strong for some transitions that the population already exhibits some decay after $\tau=1~\mu$s, which is why we normalise the signal with the end-of-pulse intensity, that is, $I_s(\tau_1)/I_r(\tau_1)$, with a probe time $\tau_1=40~\mu$s.

These operations take typically 2 minutes for the field alignment, 1 minute for the ODMR spectrum and between 1 and 10 minutes for the $T_1$ data depending on the probe time $\tau$ used. As a result, a full spectrum takes from a few hours to tens of hours to acquire. All experiments are performed at room temperature. 

\subsection{Theoretical methods}

\subsubsection{Derivation of the spin relaxation curve} \label{sec:methods:T1curve}

\begin{figure}[b]
\begin{center}
\includegraphics[width=0.3\textwidth]{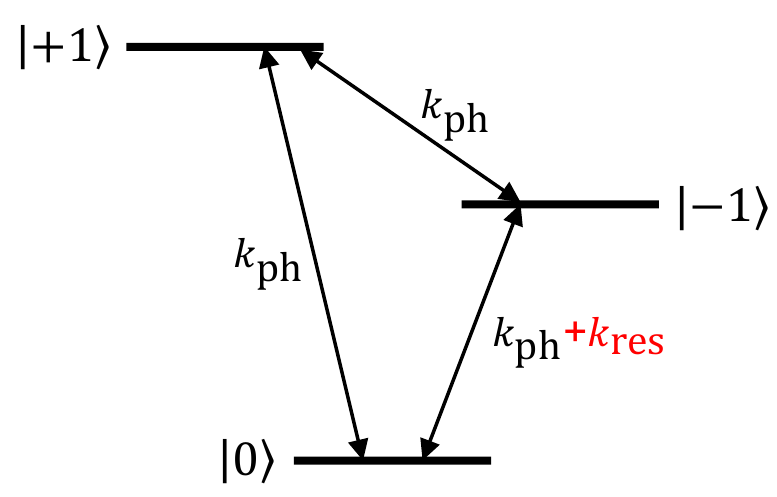}
\caption{Model used to describe the population dynamics within the NV electronic ground state, i.e., between the spin states $|0\rangle$, $|+1\rangle$ and $|-1\rangle$. The transition rate $k_{\rm ph}$ accounts for the phonon-induced relaxation while $k_{\rm res}$ corresponds to the interaction with target spins on resonance with the $|0\rangle\rightarrow|-1\rangle$ transition.} 
\label{Fig7}
\end{center}
\end{figure}

To derive the NV spin relaxation curve (Eq. (\ref{eq:T1curve})), we consider a closed three-level system composed of the three spin states of the NV ground state, $|0\rangle$, $|+1\rangle$ and $|-1\rangle$ (Fig. \ref{Fig7}). The corresponding populations are denoted $n_{0}$, $n_{+1}$ and $n_{-1}$, respectively. In the absence of interaction with target spins, the population dynamics is governed by phonon relaxation processes \cite{Jarmola2012}. In our model, this phonon-induced relaxation is accounted for through two-way transition rates between all states, with a constant rate $k_{\rm ph}$. When the resonance condition (\ref{eq:rescond}) is fulfilled, cross relaxation with the target spins provides an additional relaxation channel between $|0\rangle$ and $|-1\rangle$, with a transition rate $k_{\rm res}$. Solving the rate equations together with the closed-system condition $n_{0}(\tau)+n_{+1}(\tau)+n_{-1}(\tau)=1$ yields the populations
\begin{eqnarray} \label{eq:populations}
&n_0(\tau)=&\frac{1}{3}+\frac{1}{2}\left[\frac{1}{3}-n_{+1}(0)\right]e^{-3k_{\rm ph}\tau} \nonumber \\
& & +\frac{1}{2}\left[n_0(0)-n_{-1}(0)\right]e^{-3k_{\rm ph}\tau-2k_{\rm res}\tau} \nonumber \\
&n_{-1}(\tau)=&\frac{1}{3}+\frac{1}{2}\left[\frac{1}{3}-n_{+1}(0)\right]e^{-3k_{\rm ph}\tau}  \\
& & -\frac{1}{2}\left[n_0(0)-n_{-1}(0)\right]e^{-3k_{\rm ph}\tau-2k_{\rm res}\tau} \nonumber \\
&n_{+1}(\tau)=&\frac{1}{3}-\left[\frac{1}{3}-n_{+1}(0)\right]e^{-3k_{\rm ph}\tau}. \nonumber 
\end{eqnarray}
The PL intensity at the start of the readout laser pulse, $I_s(\tau)$, can be expressed as 
\begin{align} \label{eq:PLdef}
\begin{split}
I_s(\tau)&=I_0n_0(\tau)+I_1[n_{+1}(\tau)+n_{-1}(\tau)] \\
&=I_1+(I_0-I_1)n_0(\tau)
\end{split}
\end{align}
where $I_0$ and $I_1<I_0$ are the PL rates associated with spin states $|0\rangle$ and $|\pm1\rangle$. Inserting Eqs.~(\ref{eq:populations}) into Eq. (\ref{eq:PLdef}), the relaxation curve can be written as
\begin{equation} \label{eq:PLfunction}
I_s(\tau)=I_\infty\left[1+\frac{\cal C}{4}\left(e^{-\Gamma_{1,\rm ph}\tau}+3e^{-(\Gamma_{1,\rm ph}+\Gamma_{1,\rm res})\tau}\right)\right]
\end{equation}
where we introduced $\Gamma_{1,\rm ph}=3k_{\rm ph}$ and $\Gamma_{1,\rm res}=2k_{\rm res}$. The coefficients $I_\infty$ and ${\cal C}$ are given by  
\begin{align} \label{eq:PLcoeff}
\begin{split}
I_\infty&=\frac{I_0+2I_1}{3}\\
{\cal C}&=\frac{1-\alpha}{1+2\alpha}\left[3n_0(0)-1\right]
\end{split}
\end{align}
with $\alpha=I_1/I_0\approx0.7$ under typical experimental conditions. In obtaining Eq. (\ref{eq:PLfunction}) we assumed that the initial populations are such that $n_{+1}(0)=n_{-1}(0)$, that is, the initialisation pulse affects the populations of $|\pm1\rangle$ in the same way, as it is generally accepted \cite{Manson2006,Robledo2011}.

\subsubsection{Hamiltonian of the NV centre} \label{sec:methods:NV}

\begin{figure*}[t]
\begin{center}
\includegraphics[width=0.75\textwidth]{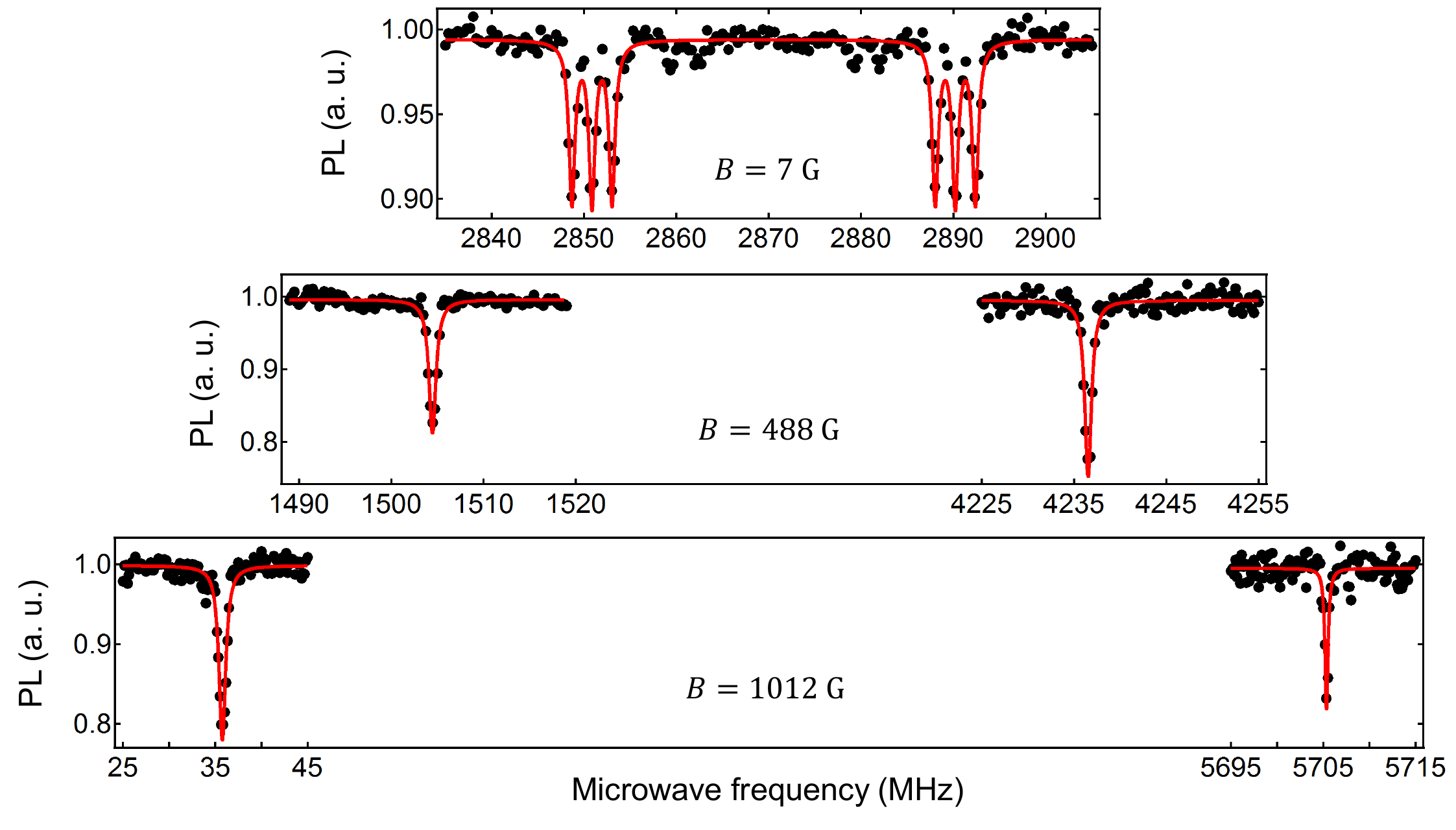}
\caption{ODMR spectra of a single NV centre recorded at different magnetic field strengths (increasing from top to bottom). Solid lines are data fitting to a sum of Lorentzian functions. The nuclear spin intrinsic to the NV centre is efficiently polarised at high fields under optical pumping, reducing the number of ODMR lines from 6 down to 2.} 
\label{Fig8}
\end{center}
\end{figure*}

The NV centre, used as the probe, comprises an electronic spin $S=1$ and a nuclear spin $I=1$ associated to the $^{14}$N nucleus. In the electronic ground state, the spin Hamiltonian can be written as
\begin{eqnarray} \label{eq:NVHam2}
\frac{{\cal H}_{\rm NV}}{h} &=& D_{\rm NV}\left(S_z^{\rm (p)}\right)^2 - \tilde{\gamma}_{\rm NV}B S_z^{\rm (p)} + a_\parallel S_z^{\rm (p)} I_z^{\rm (p)} \nonumber \\
& & +a_\perp \left(S_x^{\rm (p)} I_x^{\rm (p)} + S_y^{\rm (p)} I_y^{\rm (p)}\right) \\
& & +q\left(I_z^{\rm (p)}\right)^2-\tilde{\gamma}_N B I_z^{\rm (p)} \nonumber
\end{eqnarray}
where $\vec{S}^{\rm (p)}$ and $\vec{I}^{\rm (p)}$ are the electron and nuclear spin operators, $a_\parallel=-2.14$ MHz and $a_\perp=-2.70$ MHz are the axial and transverse hyperfine coupling parameters, and $q=-4.96$ MHz is the quadrupolar coupling parameter \cite{Fuchs2011}. The magnetic field is assumed to be aligned along $z$ with a strength $B$. In this work, both ODMR and relaxometry probe the NV electronic spin transitions from $m_S^{(\rm p)}=0$ to $m_S^{(\rm p)}=-1$ while conserving the nuclear spin projection ($\Delta m_I^{(\rm p)}=0$). Diagonalising the Hamiltonian (\ref{eq:NVHam2}) gives the transition frequencies which are, to first order in $a_\perp/(D_{\rm NV}-\tilde{\gamma}_{\rm NV}B)$,
\begin{eqnarray} \label{eq:wNVfull}
\omega_{\rm NV}\left(m_I^{(\rm p)}=+1\right) &=& D_{\rm NV}-\tilde{\gamma}_{\rm NV}B-a_\parallel+\frac{a_\perp^2}{D_{\rm NV}-\tilde{\gamma}_{\rm NV}B} \nonumber \\
\omega_{\rm NV}\left(m_I^{(\rm p)}=0\right) &=& D_{\rm NV}-\tilde{\gamma}_{\rm NV}B+\frac{2a_\perp^2}{D_{\rm NV}-\tilde{\gamma}_{\rm NV}B} \\
\omega_{\rm NV}\left(m_I^{(\rm p)}=-1\right) &=& D_{\rm NV}-\tilde{\gamma}_{\rm NV}B+a_\parallel+\frac{a_\perp^2}{D_{\rm NV}-\tilde{\gamma}_{\rm NV}B}. \nonumber
\end{eqnarray}
Similar expressions can be derived for the $m_S^{(\rm p)}=0\rightarrow+1$ transitions. These transitions would be required for a target with a zero-field splitting greater than that of the NV ($D_{\rm t} > D_{\rm NV}$). 

Fig. \ref{Fig8} shows example ODMR spectra recorded at different magnetic field strengths. At low field ($B=7$ G), the three transitions are visible for each branch. At higher fields however, only two transitions are observed. This is because the nuclear spin is polarised in the $m_I^{(\rm p)}=+1$ nuclear spin state under optical pumping due to the proximity of the ESLAC or GSLAC \cite{Jacques2009,Fuchs2011}. As a result, the NV transition relevant to this work has a frequency
\begin{equation} \label{eq:NVfreq}
\omega_{\rm NV}=D_{\rm NV}-\tilde{\gamma}_{\rm NV}B-a_\parallel
\end{equation}
where we dropped the last term in Eq. (\ref{eq:wNVfull}) as it is negligible ($<100$ kHz) in the range of field strengths used in this work to detect the P1 resonances. This is the formula used to convert the NV transition frequency $\omega_{\rm NV}$ determined from the ODMR spectrum into the magnetic field strength $B$, where the zero-field splitting $D_{\rm NV}$ is obtained from the low-field ODMR spectrum. For the NV centre used in Figs. \ref{Fig2} and \ref{Fig4}, we measured $D_{\rm NV}=2870.5(1)$ MHz.

\subsubsection{Transition frequencies of the P1 centre} \label{sec:methods:P1}

The P1 centre in diamond \cite{Loubser1978,Smith1959,Cook1966,Hanson2008}, used as our target, contains an electron spin $S=\frac{1}{2}$ associated with an unpaired electron, and a nuclear spin $I=1$ associated with the $^{14}$N nucleus. The unpaired electron is shared by the nitrogen atom and the neighbouring carbon atom. The delocalisation of the electron is accompanied by the Jahn-Teller elongation of the corresponding carbon-nitrogen bond. Because there are four equivalent neighbouring carbons around the nitrogen, the P1 centre can have four possible symmetry axes: $[111]$, which also corresponds to the NV centre's symmetry axis, or one of the non-parallel axes $[\bar{1}11]$, $[1\bar{1}1]$ and $[11\bar{1}]$. Since we wish to express the Hamiltonian of the P1 centre in the $z$-basis of the NV centre, the Hamiltonian will take two different forms for the on-axis or off-axis cases. In what follows we will detail the on-axis case, and then describe how to deduce the off-axis case. 

The Hamiltonian of a P1 centre with symmetry axis along $z$ can be written as
\begin{align} \label{eq:P1Ham}
\begin{split}
\frac{{\cal H}_{\rm P1}}{h} = & -\tilde{\gamma}_eBS_z-\tilde{\gamma}_{N}BI_z +A_\parallel S_z I_z \\
& +A_\perp\left(S_xI_x+S_yI_y\right)+QI_z^2
\end{split}   
\end{align}
where $\vec{S}$ and $\vec{I}$ are the P1 electron and nuclear spin operators (note that we use the same notations as for the NV spin operators as there is no ambiguity) and the various parameters are defined in Sec. \ref{sec:exp}. Here the magnetic field is assumed to be aligned along $z$ with a strength $B$. Diagonalising the Hamiltonian (\ref{eq:P1Ham}) gives the energy levels of the system. Retaining terms up to and including order ${\cal O}\left(\frac{A_\perp^2}{\omega_e^2}\right)$ (where $\omega_e = -\tilde{\gamma}_eB$) these energies are
\begin{align} \label{eq:EP1}
\begin{split}
\frac{E}{h}\left(\left\vert+\tfrac{1}{2},+1\right\rangle\right) &= \frac{\omega_e}{2} +  \frac{A_\parallel}{2} + Q +\omega_N\\
\frac{E}{h}\left(\left\vert+\tfrac{1}{2},0\right\rangle\right) &= \frac{\omega_e}{2} +\frac{A_\perp^2}{2\omega_e}\\
\frac{E}{h}\left(\left\vert+\tfrac{1}{2},-1\right\rangle\right) &=\frac{\omega_e}{2}  -  \frac{A_\parallel}{2}+\frac{A_\perp^2}{2\omega_e} + Q - \omega_N\\
\frac{E}{h}\left(\left\vert-\tfrac{1}{2},+1\right\rangle\right) &= -\frac{\omega_e}{2}  -  \frac{A_\parallel}{2}-\frac{A_\perp^2}{2\omega_e} + Q +\omega_N\\
\frac{E}{h}\left(\left\vert-\tfrac{1}{2},0\right\rangle\right) &= -\frac{\omega_e}{2} -\frac{A_\perp^2}{2\omega_e}\\
\frac{E}{h}\left(\left\vert-\tfrac{1}{2},-1\right\rangle\right) &= -\frac{\omega_e}{2}  +  \frac{A_\parallel}{2} + Q - \omega_N
\end{split}
\end{align}
where the ket $\vert m_S^{(\rm t)},m_I^{(\rm t)}\rangle$ indicates the state of the unperturbed P1 spin system. The states whose energy contains a term $A_\perp^2/2\omega_e$ are in fact perturbations on the electron and nuclear spin states stated since the transverse hyperfine coupling ($A_\perp$) causes state mixing. In Eq. (\ref{eq:EP1}) we introduced the Zeeman shift for the electron $\omega_e=-\tilde{\gamma}_eB$ and for the nucleus $\omega_N=-\tilde{\gamma}_NB$. 

The single-quantum EPR transitions probed in Fig. \ref{Fig2} correspond to transitions from $m_S^{(\rm t)}=+1/2$ to $m_S^{(\rm t)}=-1/2$ while conserving the nuclear spin projection ($\Delta m_I^{(\rm t)}=0$). Using Eqs. (\ref{eq:EP1}), one can express these transition frequencies as 
\begin{align} \label{eq:wEPRsingle}
\begin{split}
\omega_{\rm t}\left(\left\vert+\tfrac{1}{2},+1\right\rangle\rightarrow\left\vert-\tfrac{1}{2},+1\right\rangle\right) &= \omega_e + A_\parallel +\frac{A_\perp^2}{2\omega_e} \\
\omega_{\rm t}\left(\left\vert+\tfrac{1}{2},0\right\rangle\rightarrow\left\vert-\tfrac{1}{2},0\right\rangle\right) &= \omega_e + \frac{A_\perp^2}{\omega_e} \\
\omega_{\rm t}\left(\left\vert+\tfrac{1}{2},-1\right\rangle\rightarrow\left\vert-\tfrac{1}{2},-1\right\rangle\right) &= \omega_e - A_\parallel +\frac{A_\perp^2}{2\omega_e} 
\end{split}
\end{align}
Likewise, the frequencies of the allowed double-quantum transitions are
\begin{align} \label{eq:wEPRdouble}
\begin{split}
\omega_{\rm t}\left(\left\vert+\tfrac{1}{2},0\right\rangle\rightarrow\left\vert-\tfrac{1}{2},+1\right\rangle\right) &= \omega_e + \frac{A_\parallel}{2} +\frac{A_\perp^2}{2\omega_e}-Q+\omega_N \\
\omega_{\rm t}\left(\left\vert+\tfrac{1}{2},-1\right\rangle\rightarrow\left\vert-\tfrac{1}{2},0\right\rangle\right) &= \omega_e - \frac{A_\parallel}{2} +\frac{A_\perp^2}{2\omega_e}+Q+\omega_N.
\end{split}
\end{align}

The transition frequencies expressed in Eqs. (\ref{eq:wEPRsingle}) and (\ref{eq:wEPRdouble}) are written as a function of magnetic field strength $B$ via $\omega_e=-\tilde{\gamma}_eB$ and $\omega_N=-\tilde{\gamma}_NB$. To obtain the resonant frequencies as observed in the NV relaxometry data, one needs to solve for $B$ in the resonance condition $\vert\omega_{\rm NV}(B)|=|\omega_{\rm t}(B)\vert$ where $\omega_{\rm NV}(B)$ is the NV transition frequency as given by Eq. (\ref{eq:NVfreq}) and $\omega_{\rm t}(B)$ is one of the P1 transition frequencies given in  Eqs. (\ref{eq:wEPRsingle}) and (\ref{eq:wEPRdouble}). Once the resonant field $B_{\rm res}$ is found, one can compute the resonant frequency $\omega_{\rm NV}(B_{\rm res})$. This was done numerically to obtain the theoretical values in Table \ref{Table:EPR}. However, one can derive simple approximate expressions if one retains only terms up to order ${\cal O}\left(\frac{a_\parallel^2}{D_{\rm NV}^2}\right)$ and ${\cal O}\left(\frac{A_\perp^2}{D_{\rm NV}^2}\right)$. This leads to
\begin{align} \label{eq:wEPRsols}
\begin{split}
\omega_{\rm t}\left(\left\vert+\tfrac{1}{2},+1\right\rangle\rightarrow\left\vert-\tfrac{1}{2},+1\right\rangle\right) =& \frac{D_{\rm NV}-a_\parallel}{2}+\frac{A_\parallel}{2}+\frac{A_\perp^2}{2D_{\rm NV}} \\
\omega_{\rm t}\left(\left\vert+\tfrac{1}{2},0\right\rangle\rightarrow\left\vert-\tfrac{1}{2},0\right\rangle\right) =& \frac{D_{\rm NV}-a_\parallel}{2}+\frac{A_\perp^2}{D_{\rm NV}} \\
\omega_{\rm t}\left(\left\vert+\tfrac{1}{2},-1\right\rangle\rightarrow\left\vert-\tfrac{1}{2},-1\right\rangle\right) =& \frac{D_{\rm NV}-a_\parallel}{2}-\frac{A_\parallel}{2}+\frac{A_\perp^2}{2D_{\rm NV}} \\
\omega_{\rm t}\left(\left\vert+\tfrac{1}{2},0\right\rangle\rightarrow\left\vert-\tfrac{1}{2},+1\right\rangle\right) =& \frac{D_{\rm NV}-a_\parallel}{2}+\frac{A_\parallel}{4}-\frac{Q}{2} \\
 & +\frac{\tilde{\gamma}_N D_{\rm NV}}{4}+\frac{A_\perp^2}{2D_{\rm NV}} \\
\omega_{\rm t}\left(\left\vert+\tfrac{1}{2},-1\right\rangle\rightarrow\left\vert-\tfrac{1}{2},0\right\rangle\right) =& \frac{D_{\rm NV}-a_\parallel}{2}-\frac{A_\parallel}{4}+\frac{Q}{2} \\
 & +\frac{\tilde{\gamma}_N D_{\rm NV}}{4}+\frac{A_\perp^2}{2D_{\rm NV}} \\
\end{split}
\end{align}
where we used the approximation $\tilde{\gamma}_{\rm NV}=\tilde{\gamma}_e$.

The NMR transitions probed in Fig. \ref{Fig4} correspond to transitions that conserve the electron spin projection ($\Delta m_S^{(\rm t)}=0$). The corresponding frequencies obtained from Eqs. (\ref{eq:EP1}) are
\begin{align} \label{eq:wNMR}
\begin{split}
\omega_{\rm t}\left(\left\vert+\tfrac{1}{2},+1\right\rangle\leftrightarrow\left\vert+\tfrac{1}{2},0\right\rangle\right) &= \frac{A_\parallel}{2} -\frac{A_\perp^2}{2\omega_e}+Q-\omega_N \\
\omega_{\rm t}\left(\left\vert+\tfrac{1}{2},-1\right\rangle\leftrightarrow\left\vert+\tfrac{1}{2},0\right\rangle\right) &= \frac{A_\parallel}{2} -Q-\omega_N \\
\omega_{\rm t}\left(\left\vert-\tfrac{1}{2},+1\right\rangle\leftrightarrow\left\vert-\tfrac{1}{2},0\right\rangle\right) &= \frac{A_\parallel}{2} -Q+\omega_N \\
\omega_{\rm t}\left(\left\vert-\tfrac{1}{2},-1\right\rangle\leftrightarrow\left\vert-\tfrac{1}{2},0\right\rangle\right) &= \frac{A_\parallel}{2} +\frac{A_\perp^2}{2\omega_e}+Q+\omega_N. \\
\end{split}
\end{align}
Again, the frequencies at resonance with the NV spin can be computed numerically, the resulting values being given in Table \ref{Table:NMR}.  Because these frequencies depend weakly on $B$ and are much smaller than $D_{\rm NV}$, the resonant field is approximately $B_{\rm res}\approx D_{\rm NV}/\tilde{\gamma}_{\rm NV}\approx1024$ G and the Zeeman shifts $\omega_e=-\tilde{\gamma}_eB_{\rm res}$ and $\omega_N=-\tilde{\gamma}_NB_{\rm res}$ can be considered as constants. The corresponding terms in Eqs. (\ref{eq:wNMR}) are $\frac{A_\perp^2}{2\omega_e}\approx1.15$ MHz (in fact, 1.173 MHz at 1006 G, 1.133 MHz at 1042 G) and $\omega_N\approx0.315$ MHz (0.309 MHz at 1006 G, 0.321 MHz at 1042 G).

For an off-axis P1 centre, the magnetic field is now forming an angle with the intrinsic quantisation axis of the P1 centre. To express the Hamiltonian in the same $z$-basis where $z$ is the direction of the magnetic field, one must apply a rotation of the spin operators. This leads to a Hamiltonian of the same form \cite{Knowles2013,Hall2015}
\begin{align} \label{eq:P1HamOff}
\begin{split}
\frac{{\cal H}'_{\rm P1}}{h} = & -\tilde{\gamma}_eBS_z-\tilde{\gamma}_{N}BI_z +A_\parallel' S_z I_z \\
& +A_\perp'(S_xI_x+S_yI_y)+Q'I_z^2
\end{split}   
\end{align}
where the apparent hyperfine and quadrupole coupling parameters are modified according to $A_\parallel'=\frac{1}{9}(A_\parallel+8A_\perp)$, $A_\perp'=\frac{1}{9}(4A_\parallel+5A_\perp)$ and $Q'=Q$. Therefore the expressions obtained for the on-axis case (Eqs. (\ref{eq:wEPRsols}) and (\ref{eq:wNMR})) are still valid upon using these modified parameters. The numerically computed resonant transition frequencies are given in Tables \ref{Table:EPR} and \ref{Table:NMR}.

\subsubsection{$T_1$-EPR/NMR on a single spin-1/2} \label{sec:methods:GammaInt}

We now turn to the calculation of the relaxation rate of the NV probe spin on resonance with a target spin system. In this section, we consider the simple case of a single spin-1/2 as the target, which is treated using a fully quantum mechanical approach identical to that employed in Ref. \cite{Hall2015}. In Sec. \ref{sec:methods:P1strength}, we will describe an alternative method well suited to describe more complicated spin systems, which we will apply to the P1 centre. 

We seek here to calculate the population dynamics of a system composed of the NV spin and a single spin-1/2 target with gyromagnetic ratio $\tilde{\gamma}_{\rm t}$. The Hamiltonian of the coupled NV-target system is
\begin{equation}
{\cal H}={\cal H}_{\rm NV}+{\cal H}_{\rm t}+{\cal H}_{\rm int}
\end{equation}
where ${\cal H}_{\rm NV}$ is the Hamiltonian of the NV spin, ${\cal H}_{\rm t}$ is that of the target spin and ${\cal H}_{\rm int}$ is the magnetic dipole-dipole interaction between the two spins. We restrict the NV spin to the $\{m_S^{(\rm p)}=0,m_S^{(\rm p)}=-1\}$ subset and neglect hyperfine interaction with the NV centre's nitrogen nuclear spin. The magnetic field of strength $B$ is applied along the NV centre's symmetry axis, which defines the $z$ direction. The Hamiltonian of the NV and target spins are therefore simply 
\begin{eqnarray} 
{\cal H}_{\rm NV}/h &=& D_{\rm NV}\left(S_z^{(\rm p)}\right)^2 - \tilde{\gamma}_{\rm NV}B S_z^{(\rm p)} \label{eq:NVHam3}\\
{\cal H}_{\rm t}/h &=& -\tilde{\gamma}_{\rm t} B I_z^{(\rm t)}\label{eq:TargHam}
\end{eqnarray}
where $\vec{I}^{(\rm t)}$ denotes the spin operator of the target (which can be an electronic or nuclear spin), and the superscripts (p) and (t) on the operators indicate whether it refers to the probe or target.

The dipole-dipole interaction is
\begin{equation} \label{eq:DDHam}
{\cal H}_{\rm int}=-\frac{\mu_0\tilde{\gamma}_{\rm NV}\tilde{\gamma}_{\rm t} h^2}{4\pi r^3}\left[\frac{3}{r^2}\left(\vec{S}^{(\rm p)}\cdot\vec{r}\right)\left(\vec{I}^{(\rm t)}\cdot\vec{r}\right)-\vec{S}^{(\rm p)}\cdot\vec{I}^{(\rm t)}\right]
\end{equation}
where $\vec{r}$ is the vector joining the NV to the target and $r=|\vec{r}|$. In the $\{|0,+1/2\rangle,|0,-1/2\rangle,|-1,+1/2\rangle,|-1,-1/2\rangle\}$ basis using the notation $\vert m_S^{(\rm p)},m_I^{(\rm t)}\rangle$, the total Hamiltonian is expressed as
\begin{widetext}
\begin{equation} \label{eq:Htotmat}
\frac{\cal H}{h} = 
\begin{pmatrix}
-\frac{\tilde{\gamma}_{\rm t} B}{2} & 0 & \frac{H_{\rm int,13}}{h} & \frac{H_{\rm int,14}}{h} \\ 0 & \frac{\tilde{\gamma}_{\rm t} B}{2} & \frac{H_{\rm int,23}}{h} & \frac{H_{\rm int,24}}{h} \\ \frac{H_{\rm int,31}}{h} & \frac{H_{\rm int,32}}{h} & D_{\rm NV}+\tilde{\gamma}_{NV} B -\frac{\tilde{\gamma}_{\rm t} B}{2} + \frac{H_{\rm int,33}}{h} & \frac{H_{\rm int,34}}{h} \\ \frac{H_{\rm int,41}}{h} & \frac{H_{\rm int,42}}{h} & \frac{H_{\rm int,43}}{h} & D_{\rm NV}+\tilde{\gamma}_{NV} B +\frac{\tilde{\gamma}_{\rm t} B}{2} + \frac{H_{\rm int,44}}{h}
\end{pmatrix}
\end{equation}
\end{widetext}
where $\{H_{\rm int,ij}\}$ are the matrix elements of ${\cal H}_{\rm int}$. Denoting $\{H_{ij}\}$ as the matrix elements of ${\cal H}$, the two resonances occur when $H_{11} = H_{44}$ and $H_{22} = H_{33}$, which yields
\begin{align} \label{eq:ResExact}
\begin{split}
-\frac{\tilde{\gamma}_{\rm t} B}{2} &=  D_{NV}+\tilde{\gamma}_{NV} B +\frac{\tilde{\gamma}_{\rm t} B}{2} + \frac{H_{\rm int,44}}{h}\\
\frac{\tilde{\gamma}_{\rm t} B}{2} &=  D_{NV}+\tilde{\gamma}_{NV} B -\frac{\tilde{\gamma}_{\rm t} B}{2} + \frac{H_{\rm int,33}}{h}.
\end{split}
\end{align}
These two resonances occur before and after the GSLAC if $\tilde{\gamma}_{\rm t}<0$, respectively, and after and before the GSLAC if $\tilde{\gamma}_{\rm t}>0$. Note that for weak dipolar coupling the terms $H_{\rm int,33}$ and $H_{\rm int,44}$ in Eqs. (\ref{eq:ResExact}) can be neglected leading to Eq. (\ref{eq:Bres}) for the resonant fields $B_{\rm res}^\pm$.

Using the Lindblad equation, the rate of change of the density matrix $\rho(t)$ is 
\begin{widetext}
\begin{align}
\begin{split}
\dot{\rho}\left(t\right) =  -\frac{i}{\hbar}\left[{\cal H}\rho\left(t\right) - \rho\left(t\right){\cal H}\right] 
& + 2\Gamma_2^{\rm (p)} \left[S_z^{\rm (p)}\rho\left(t\right)S_z^{\rm (p)} - \frac{1}{2}\rho\left(t\right)S_z^{\rm (p)}S_z^{(p)} - \frac{1}{2} S_z^{\rm (p)}S_z^{\rm (p)}\rho\left(t\right)\right] \\ 
& + 2\Gamma_2^{\rm (t)} \left[I_z^{\rm (t)}\rho\left(t\right)I_z^{\rm (t)} - \frac{1}{2}\rho\left(t\right)I_z^{\rm (t)}I_z^{\rm (t)} - \frac{1}{2} I_z^{\rm (t)}I_z^{\rm (t)}\rho\left(t\right)\right]
\end{split}
\label{eq:Lindblad}
\end{align}
\end{widetext}

where $\Gamma_2^{\rm (p)}$ and $\Gamma_2^{\rm (t)}$ are the dephasing rates of the NV and target spin, respectively.

When on resonance, only those terms in the resonant states need to be considered. Hence we obtain a system of first order differential equations involving only $\rho_{11}\left(t\right)$, $\rho_{14}\left(t\right)$, $\rho_{41}\left(t\right)$ and $\rho_{44}\left(t\right)$ for the resonance where $H_{11} = H_{44}$, and a system involving only $\rho_{22}\left(t\right)$, $\rho_{23}\left(t\right)$, $\rho_{32}\left(t\right)$ and $\rho_{33}\left(t\right)$ for the resonance where $H_{22} = H_{33}$. From these, and assuming the NV is fully initialised while the target spin starts in an arbitrary mixture, i.e. $\rho_{11}(0)+\rho_{22}(0)=1$, we can generate a single third order differential equation for each resonance,
\begin{widetext}
\begin{equation}
\frac{d^3\rho_{AA}\left(t\right)}{dt^3} = -2\Gamma_{2}\frac{d^2\rho_{AA}\left(t\right)}{dt^2}-\left[\omega_{\rm int,AB}^2+\Gamma_{2}^2\right]\frac{d\rho_{AA}\left(t\right)}{dt}-\Gamma_{2}\omega_{\rm int,AB}^2\rho_{AA}\left(t\right)+\frac{\Gamma_{2}}{2}\omega_{\rm int,AB}^2\rho_{AA}\left(0\right)
\end{equation}
\end{widetext}
where we introduced the effective interaction strength $\omega_{\rm int,AB}=2\frac{H_{\rm int,AB}}{\hbar}$, the total dephasing rate $\Gamma_{2}=\Gamma_{2}^{\rm (p)}+\Gamma_{2}^{\rm (t)}$ and $A,B = 1,4$ or $2,3$ depending on which resonance is being interrogated. With the initial conditions
\begin{align}
\begin{split}
\frac{d\rho_{AA}\left(0\right)}{dt} &= 0\\
\frac{d^2\rho_{AA}\left(0\right)}{dt^2} &= -\frac{\omega_{\rm int,AB}^2}{2}\rho_{AA}\left(0\right)
\end{split}
\end{align}
we find the solution
\begin{widetext}
\begin{equation} \label{eq:fullsolution}
\rho_{AA}(t) = \frac{\rho_{AA}(0)}{2}+\frac{\rho_{AA}(0)}{2}e^{-\frac{\Gamma_2}{2}t}\left[\cosh\left(\frac{t}{2}\sqrt{\Gamma_2^2-4 \omega_{\rm int,AB}^2}\right)+\frac{\Gamma_2}{\sqrt{\Gamma_2^2-4\omega_{\rm int,AB}^2}}\sinh\left(\frac{t}{2}\sqrt{\Gamma_2^2-4 \omega_{\rm int,AB}^2}\right)\right] .
\end{equation}
\end{widetext}
In the regime of weak dephasing ($\hbar\Gamma_2\ll H_{\rm int,AB}$), we obtain oscillations corresponding to a coherent energy transfer back and forth between the two spins. In most practical cases however, the coupling strength is much weaker than the dephasing ($\hbar\Gamma_2\gg H_{\rm int,AB}$). In this weak coupling regime, Eq. (\ref{eq:fullsolution}) simplifies into
\begin{align}
\rho_{AA}(t) \approx \frac{\rho_{AA}(0)}{2}+\frac{\rho_{AA}(0)}{2}e^{-\frac{\omega_{\rm int,AB}^2}{\Gamma_2}t}
\label{eq:decorate}
\end{align}
from which we identify the relaxation rate
\begin{equation}
\Gamma_{1,\rm res}=\frac{\omega_{\rm int,AB}^2}{\Gamma_{2}}=\frac{4}{\Gamma_{2}} \left(\frac{H_{\rm int,AB}}{\hbar}\right)^2.
\end{equation}
This gives, for the two resonances $H_{11} = H_{44}$ (labelled `+') and $H_{22} = H_{33}$ (labelled `-'),
\begin{align}
\begin{split}
\Gamma_{1,\rm res}^+ &= \frac{1}{\Gamma_{2}}\left(\frac{\mu_0\tilde{\gamma}_{\rm NV}\tilde{\gamma}_t h}{2\sqrt{2}} \right)^2 \left(\frac{3\sin^2\theta}{r^3} \right)^2 \\
\Gamma_{1,\rm res}^- &= \frac{1}{\Gamma_{2}}\left(\frac{\mu_0\tilde{\gamma}_{\rm NV}\tilde{\gamma}_t h}{2\sqrt{2}} \right)^2 \left(\frac{3\sin^2\theta-2}{r^3} \right)^2.
\end{split}
\label{eqn:ResPoints}
\end{align}
This corresponds to Eq. (\ref{eq:GammaInt+-}), which was discussed in the context of a nuclear spin as the target. For an electronic target spin ($\tilde{\gamma}_{\rm t}=\tilde{\gamma}_e\approx\tilde{\gamma}_{\rm NV}$), there is only one resonance at $B_{\rm res}^+\approx512$ G because the second resonance corresponds to a field $B_{\rm res}^-=\infty$. The decay rate associated with the $B_{\rm res}^+$ resonance was given in Eq. (\ref{eq:GintP1}).

We note that $\Gamma_{1,\rm res}$ does not depend on the initial state of the target spin, i.e. on the value $\rho_{AA}(0)$. However, the latter affects the relative contrast of the decay as measured via the PL. Indeed, the PL intensity can be expressed using Eq. (\ref{eq:PLdef}) as
\begin{eqnarray}  \label{eq:PLsinglespin}
I_s(\tau)&=& I_1+(I_0-I_1)\left[\rho_{11}(\tau)+\rho_{22}(\tau)\right] \nonumber \\
&=& I_0+I_0{\cal C}\left[e^{-\Gamma_{1,\rm res}\tau}-1\right]
\end{eqnarray}
where the contrast is given by
\begin{equation} 
{\cal C}=\frac{I_0-I_1}{2I_0}\rho_{AA}(0).
\end{equation}
The contrast is maximum if the target spin is initialised in the resonant state ($\rho_{AA}(0)=1$), and is null if the target spin is initialised in the other (non-resonant) state ($\rho_{AA}(0)=0$). 

In general, the target spin is initially in a mixed state ($\rho_{AA}(0)=1/2$) since the thermal energy greatly exceeds the Zeeman energy, i.e. $k_BT\gg \hbar\gamma_{\rm t}B$. However, in the presence of multiple target spins (e.g., multiple P1 centres as in our experiment), the probability of finding at least one target spin on resonance with the NV is close to unity, which implies that the contrast is maximum, i.e. ${\cal C}=(I_0-I_1)/2I_0$. The effective relaxation rate, $\Gamma_{1,\rm eff}$, is then simply a sum of the relaxation rates induced by each on-resonance target spin. Under this assumption, Eq. (\ref{eq:PLsinglespin}) is found to match Eq. (\ref{eq:PLfunction}) when one sets $\Gamma_{1, \rm ph}=0$ (no phonon relaxation) and $n_0(0)=1$ (NV electron spin fully initialised in $m_S^{\rm (p)}=0$). Eq. (\ref{eq:PLfunction}) is therefore a generalisation of Eq. (\ref{eq:PLsinglespin}) which includes phonon relaxation and non-perfect NV initialisation, and is valid in the presence of multiple target spins.

\subsubsection{$T_1$-EPR/NMR on a P1 centre} \label{sec:methods:P1strength}

In the previous section we used a fully quantum mechanical approach to treat the case where the NV spin interacts with a single spin-1/2 target. To treat the more complex case of the P1 centre, which comprises a spin-1/2 electron and a spin-1 nucleus, we employ a semi-classical approach where the NV quantum dynamics is calculated under the classical magnetic field generated by the target spin system. In Ref. \cite{Hall2015}, it has been shown that the two approaches give identical results for the case of a single spin-1/2 target. According to the semi-classical approach, the NV relaxation rate at a given background magnetic field $B$ can be expressed as \cite{Hall2015}
\begin{equation} \label{GeneralNVT1}
\Gamma_1(B)=\int 4b^2(\omega) \frac{\Gamma_2^{\rm (p)}}{[\Gamma_2^{\rm (p)}]^2+\left[\omega_\mathrm{NV}(B)-\omega\right]^2}P_B(\omega)\mathrm{d}\omega
\end{equation}
where $\Gamma_2^{\rm (p)}$ is the dephasing rate of the NV spin, $\omega_\mathrm{NV}$ is the transition frequency of the NV spin, $P_B(\omega)$ is the normalised distribution of transition frequencies of the target spin system (i.e., the magnetic spectrum of the environment) at field $B$, and $b$ is the mutual coupling strength between probe and target. The task of determining the relaxation rate of the NV spin thus reduces to computing the associated coupling strengths, $b$, and frequency spectra, $P_B(\omega)$, of the environment. This is achieved via examination of the Hamiltonian components associated with the NV-target interaction, and self-interactions within the target system, respectively.

If the target is a single spin-1/2 with gyromagnetic ratio $\tilde{\gamma}_{\rm t}$, the coupling strength is obtained from the term of the dipole-dipole interaction ${\cal H}_{\rm int}$ that corresponds to the resonance condition (see Eqs. (\ref{eq:DDHam}) and (\ref{eq:Htotmat})), which gives 
\begin{eqnarray} \label{eq:bpm}
b_\pm = \left(\frac{\mu_0\tilde{\gamma}_{\rm NV}\tilde{\gamma}_{\rm t} h}{4\sqrt{2}}\right)\left(\frac{3\sin^2\theta-1\pm1}{r^3}\right)
\end{eqnarray}
where the sign $\pm$ refers to the two possible resonances. For the P1 centre, Eq. (\ref{eq:bpm}) can be used to model the single-quantum EPR transitions (only $b_+$ in that case, with $\tilde{\gamma}_{\rm t}=\tilde{\gamma}_e$), the hyperfine-enhanced NMR transitions (with $\tilde{\gamma}_{\rm t}=\tilde{\gamma}_e$), and the direct NMR transitions (with $\tilde{\gamma}_{\rm t}=\tilde{\gamma}_N$). For the double-quantum EPR transitions, the relevant term in the dipole-dipole interaction leads to an interaction strength \cite{Hall2015} 
\begin{eqnarray} \label{eq:bdouble}
b_{\rm double} = \left(\frac{\mu_0\tilde{\gamma}_{\rm NV}\tilde{\gamma}_{\rm t} h}{8\sqrt{2}}\right)\left(\frac{3\sin 2\theta}{r^3}\right).
 \end{eqnarray}

To determine the dynamic behaviour of the P1 environment, we compute the autocorrelation functions associated with the field components of the target spin system. Interactions between target spins may be modelled by damping these autocorrelation functions with a decaying exponential, $\exp(-\Gamma_2^{\rm (t)}t)$, to describe their relaxation due to mutual flip-flop processes with corresponding relaxation rate $\Gamma_2^{\rm (t)}$.
The propagator associated with the target spin system is given by
\begin{eqnarray}
\mathcal{U}_\mathrm{t}(t) &=& \exp(-i \mathcal{\mathcal{H}}_\mathrm{t}t)
\end{eqnarray}
where we will take ${\cal H}_{\rm t}={\cal H}_{\rm P1}$ as given in Eq. (\ref{eq:P1Ham}) to treat the P1 problem. \\

{\bf Single-quantum EPR transitions.} 
In the case of single-quantum transitions of the P1 centre, the relaxation of the NV spin is caused by its coupling to the lateral components of the P1 spin. Thus, we compute the autocorrelation function associated with the lateral dynamics of the P1 spin,
\begin{eqnarray} \label{eq:AutocorrSingle}
\left\langle S_x(t)S_x(0)\right\rangle_\mathrm{single} &=& e^{-\Gamma_2^{\rm (t)}t}~\mathrm{Tr}\left\{\mathcal{U}_\mathrm{t}(t)S_x\mathcal{U}^\dag_\mathrm{t}(t)S_x\right\} \nonumber \\
 &=& e^{-\Gamma_2^{\rm (t)}t}\sum_{\{\omega_{{\rm t},i}\}}\cos \left(\omega_{{\rm t},i}t \right)
\end{eqnarray}
where the sum runs over the three Larmor precession frequencies $\{\omega_{{\rm t},i}\}$ corresponding to the single-quantum EPR transitions, as given in Eq. (\ref{eq:wEPRsingle}), assuming an on-axis P1 centre: one for each possible nuclear spin state ($m_I^{\rm (t)}=0,\pm1$). The corresponding spectrum may then be found by computing the Fourier transform of the autocorrelation function, which gives
\begin{eqnarray} \label{eq:Ssingle}
P_\mathrm{single}(\omega) &=&  \sum_{\{\omega_{{\rm t},i}\}}\frac{\Gamma_2^{\rm (t)}}{(\Gamma_2^{\rm (t)})^2+(\omega-\omega_{{\rm t},i})^2}.
\end{eqnarray}
Inserting Eqs. (\ref{eq:bpm}) and (\ref{eq:Ssingle}) into Eq. (\ref{GeneralNVT1}) gives the relaxation rate as a function $B$ about the single-quantum EPR transitions,
\begin{widetext}
\begin{eqnarray} \label{eq:specsingle}
\Gamma_{1}^{\rm EPR}(B)=\left(\frac{\mu_0\tilde{\gamma}_{\rm NV}\tilde{\gamma}_e h}{2\sqrt{2}}\right)^2\left(\frac{3\sin^2\theta}{r^3}\right)^2 \sum_{\{\omega_{{\rm t},i}\}} \frac{\Gamma_2^{\rm (p)}+\Gamma_2^{\rm (t)}}{[\Gamma_2^{\rm (p)}+\Gamma_2^{\rm (t)}]^2+\left[\omega_\mathrm{NV}(B)-\omega_{{\rm t},i}(B)\right]^2}.
\end{eqnarray} 
\end{widetext}
This spectrum comprises three Lorentzian peaks corresponding to the three possible P1 nuclear spin states ($m_I^{\rm (t)}=0,\pm1$). The amplitude of each peak, that is, the NV relaxation rate on resonance with a single-quantum EPR transition of the P1, matches that obtained using the fully quantum mechanical approach for a single spin-1/2 (see Eq. (\ref{eqn:ResPoints})). Moreover, Eq. (\ref{eq:specsingle}) shows that the line width of each resonance is governed by the total dephasing rate $\Gamma_{2}=\Gamma_{2}^{(\rm p)}+\Gamma_{2}^{(\rm t)}$. \\  

{\bf Double-quantum EPR transitions.} 
Similarly, for the double-quantum EPR transitions of the P1 centre, the relaxation of the NV spin is caused by its coupling to the axial components of the P1 spin. Thus we compute,
\begin{widetext}
\begin{eqnarray}
\left\langle {S}_z(t){S}_z(0)\right\rangle_\mathrm{double} &=& e^{-\Gamma_1^{\rm (t)}t}\mathrm{Tr}\left\{\mathcal{U}_\mathrm{T}(t){S}_z\mathcal{U}^\dag_\mathrm{T}(t){S}_z\right\} \nonumber \\
&\approx & e^{-\Gamma_1^{\rm (t)}t}\left[ 1-\sum_{\{\omega_{{\rm t},j}\}}4\left(\frac{A_\perp}{\omega_e}\right)^2\sin ^2\left(\frac{\omega_{{\rm t},j}t}{2}\right)\right] 
\end{eqnarray}
\end{widetext}
where $\{\omega_{{\rm t},j}\}$ are the two frequencies corresponding to the double-quantum EPR transitions, as given in Eq. (\ref{eq:wEPRdouble}), and we retained terms up to order ${\cal O}\left(\frac{A_\perp^2}{\omega_e^2}\right)$ in the prefactor. Here the damping factor corresponds to longitudinal relaxation since it applies to the $z$ spin component, with a decay rate denoted as $\Gamma_1^{\rm (t)}$. Like before, computing the Fourier transform gives the associated spectrum, $P_\mathrm{double}(\omega)$. Inserting $P_\mathrm{double}(\omega)$ and Eq. (\ref{eq:bdouble}) into Eq. (\ref{GeneralNVT1}) gives the field-dependent relaxation rate about the double-quantum EPR transitions,
\begin{widetext}
\begin{eqnarray} \label{eq:specdouble}
\Gamma_{1}^{\rm EPR,double}(B)=\left(\frac{\mu_0\tilde{\gamma}_{\rm NV}\tilde{\gamma}_e h}{2\sqrt{2}}\right)^2\left(\frac{3\sin 2\theta}{r^3}\right)^2 \left(\frac{A_\perp}{\omega_e(B)}\right)^2 \sum_{\{\omega_{{\rm t},j}\}} \frac{\Gamma_2^{\rm (p)}+\Gamma_1^{\rm (t)}}{[\Gamma_2^{\rm (p)}+\Gamma_1^{\rm (t)}]^2+\left[\omega_\mathrm{NV}(B)-\omega_{{\rm t},j}(B)\right]^2}.
\end{eqnarray} 
\end{widetext}
The on-resonance relaxation rate for the two transitions is given in Eq. (\ref{eq:GintP1dbl}), where we defined the total dephasing rate of the interacting system as $\Gamma_2=\Gamma_2^{\rm (p)}+\Gamma_1^{\rm (t)}$. Compared with the single-quantum transitions, the relaxation rate is further damped by a factor of order $\sim A_\perp^2/\omega_e^2$. This is representative of the fact that at the point of resonance for those double-quantum transitions, the magnetisation exchange between the P1 electron and nuclear spin is not an energy-conserving process and is thus less likely to occur.  \\  

{\bf Hyperfine-enhanced NMR transitions.} 
We now turn to the description of the resonance features observed near the GSLAC of the NV spin, which occurs at $\approx 1024$ G. These features arise from two effects: the low-frequency components of the P1 electron EPR spectrum, as measured via the NV-P1 electron coupling; and the NMR spectrum of the P1 nuclear spin. In what follows, we discuss the origin of these signals, and show that it is only the former that produces an appreciable signal. These results demonstrate that electron-mediated enhancement of NMR signals is a viable mechanism for vastly improved sensing of nuclear magnetic resonance.

In determining the low frequency components of the P1 EPR spectrum near 1024 G, we proceed as above but retain terms of order ${\cal O}\left(\frac{A_\perp^2}{\omega_e^2}\right)$, where $\omega_e\approx 2.87$ GHz at 1024 G is the P1 electron Larmor frequency. Furthermore, we ignore terms of frequency near $\omega_e$, since these are two high to resonate with the NV frequency at this field. The relevant components of the autocorrelation function are given by
\begin{equation} \label{eq:SxSxAss}
\left\langle {S}_x(t){S}_x(0)\right\rangle_\mathrm{ass} = e^{-\Gamma_2^{'\rm (t)}t}\sum_{\{\omega_{{\rm t},k}\}}\left(\frac{A_\perp}{\omega _{e}}\right)^2\cos\left(\frac{\omega_{{\rm t},k}t}{2}\right)
\end{equation}
where $\{\omega_{{\rm t},k}\}$ are the four frequencies corresponding to the NMR transitions transitions, as given in Eq. (\ref{eq:wNMR}), and we discarded the higher order terms in $\frac{A_\parallel}{\omega_e}$ and $\frac{A_\perp}{\omega_e}$ in the prefactor. Note that although Eq. (\ref{eq:SxSxAss}) refers to the autocorrelation fonction of the P1 electron spin, the dephasing rate here, $\Gamma_2^{'\rm (t)}$, is that of the P1 nuclear spin. This is because the NMR frequencies $\{\omega_{{\rm t},k}\}$ do not depend (at first order) on the P1 electron Larmor frequency, $\omega _{e}$, and therefore are not affected by the associated fluctuations. Inserting the associated spectrum $P_\mathrm{ass}(\omega)$ and Eq. (\ref{eq:bpm}) into Eq. (\ref{GeneralNVT1}) gives the field-dependent relaxation rate about these hyperfine-enhanced NMR transitions,
\begin{widetext}
\begin{eqnarray} \label{eq:specNMRass}
\Gamma_{1}^{\rm NMR,hyp}(B)=\left(\frac{\mu_0\tilde{\gamma}_{\rm NV}\tilde{\gamma}_e h}{2\sqrt{2}}\right)^2\left(\frac{3\sin^2\theta-1\pm1}{r^3}\right)^2 \left(\frac{A_\perp}{\omega_e(B)}\right)^2 \sum_{\{\omega_{{\rm t},k}\}} \frac{\Gamma_2^{\rm (p)}+\Gamma_2^{'\rm (t)}}{[\Gamma_2^{\rm (p)}+\Gamma_2^{'\rm (t)}]^2+\left[\omega_\mathrm{NV}(B)-\omega_{{\rm t},k}(B)\right]^2}.
\end{eqnarray} 
\end{widetext}
The on-resonance relaxation rates of the two families of transitions are given in Eq. (\ref{eq:GintP1NMR}), where we defined the total dephasing rate of the interacting system as $\Gamma_2=\Gamma_2^{\rm (p)}+\Gamma_2^{'\rm (t)}$. In comparison with the single-quantum EPR transitions, the hyperfine-enhanced NMR transitions result in NV relaxation rates that are suppressed to an order of $\left(A_\perp/\omega_e\right)^2$.  \\

{\bf Direct NMR transitions.} 
We may apply the same approach as above to calculate the NMR spectrum associated with the direct coupling between the NV spin and the nuclear spin of the P1.
The autocorrelation function is given by 
\begin{eqnarray}
\left\langle {I}_x(t){I}_x(0)\right\rangle_\mathrm{direct} &=& e^{-\Gamma_2^{'\rm (t)}t}\sum_{\{\omega_{{\rm t},k}\}}\cos \left(\omega_{{\rm t},k}t \right)
\end{eqnarray}
where $\Gamma_2^{'\rm (t)}$ is the dephasing rate of the P1 nuclear spin.
It is readily apparent that these dynamics are not suppressed like those in the hyperfine-enhanced case. Despite this, the resulting effect on the NV relaxation of the direct NMR transitions is much lower than those of the hyperfine-enhanced NMR, due to the differences in coupling to the NV spin. Inserting the associated spectrum $P_\mathrm{direct}(\omega)$ and Eq. (\ref{eq:bpm}) into Eq. (\ref{GeneralNVT1}) gives the field-dependent relaxation rate about these direct NMR transitions,
\begin{widetext}
\begin{eqnarray} \label{eq:specNMRdir}
\Gamma_{1}^{\rm NMR,direct}(B)=\left(\frac{\mu_0\tilde{\gamma}_{\rm NV}\tilde{\gamma}_N h}{2\sqrt{2}}\right)^2\left(\frac{3\sin^2\theta-1\pm1}{r^3}\right)^2 \sum_{\{\omega_{{\rm t},k}\}} \frac{\Gamma_2^{\rm (p)}+\Gamma_2^{'\rm (t)}}{[\Gamma_2^{\rm (p)}+\Gamma_2^{'\rm (t)}]^2+\left[\omega_\mathrm{NV}(B)-\omega_{{\rm t},k}(B)\right]^2}.
\end{eqnarray} 
\end{widetext}

{\bf Summary.} 
For all of the transitions, the NV relaxation rate on resonance has the form
\begin{align} \label{eq:GammaGeneral}
\Gamma_{1,\rm res} &= \frac{1}{\Gamma_{2}}\left(\frac{\mu_0\tilde{\gamma}_{\rm NV}h}{2\sqrt{2}}\right)^2 \left(\tilde{\gamma}_{t}\right)^2 {\cal A}\Theta\left(r,\theta\right) 
\end{align}
where ${\cal A}$ comes from the coefficients of the autocorrelation function's Fourier transform, $\Theta(r,\theta)$ comes from the relevant term in the dipole-dipole interaction, $\Gamma_{2}$ includes the dephasing of the NV probe and the relevant dephasing of the P1 spin system, and $\tilde{\gamma}_{t}$ is the gyromagnetic ratio of the relevant P1 spin. Expressions of ${\cal A}$, $\Theta(r,\theta)$, $\Gamma_{2}$ and $\tilde{\gamma}_{t}$ are given in Table \ref{tab:P1strength} for the NV-P1 resonances, which are of four types: single-quantum EPR transitions, double-quantum EPR transitions, hyperfine-enhanced NMR transitions and direct NMR transitions. 

\begin{table*}[t]  
\begin{tabular}{|c|c|c|c|c|c|}
\hline
Transition type & Notation for $\Gamma_{1,\rm res}$ & $\Gamma_{2}$ & $\left(\tilde{\gamma}_{t}\right)^2$ & ${\cal A}$ & $\Theta(r,\theta)$ \\
\hline
single-quantum EPR & $\Gamma_{1,\rm res}^{\rm EPR}$ & $\Gamma_{2}^{(\rm p)}+\Gamma_{2}^{(\rm t)}$ & $\left(\tilde{\gamma}_e\right)^2$ & 1 & $\left(\frac{3 \sin^2\theta}{r^3}\right)^2$\\
\hline
double-quantum EPR & $\Gamma_{1,\rm res}^{\rm EPR,double}$ & $\Gamma_{2}^{(\rm p)}+\Gamma_{1}^{(\rm t)}$ & $\left(\tilde{\gamma}_e\right)^2$ & $\left(\frac{A_\perp}{\omega_e}\right)^2$ & $\left(\frac{3 \sin2\theta}{r^3}\right)^2$ \\
\hline
hyperfine-enhanced NMR & $\Gamma_{1,\rm res\pm}^{\rm NMR,hyp}$ & $\Gamma_{2}^{(\rm p)}+\Gamma_{2}^{'(\rm t)}$ & $\left(\tilde{\gamma}_e\right)^2$ & $\left(\frac{A_\perp}{\omega'_e}\right)^2$ & $\left(\frac{3 \sin^2\theta-1\pm1}{r^3}\right)^2$ \\
\hline
direct NMR & $\Gamma_{{1,\rm res}\pm}^{\rm NMR}$ & $\Gamma_{2}^{(\rm p)}+\Gamma_{2}^{'(\rm t)}$ & $\left(\tilde{\gamma}_N\right)^2$ & 1 & $\left(\frac{3 \sin^2\theta-1\pm1}{r^3}\right)^2$ \\
\hline
\end{tabular}
\caption{Expressions of the different terms that compose the NV relaxation rate on resonance with a target spin transition, according to Eq. (\ref{eq:GammaGeneral}), for the four situations considered in this work. The second column indicates the notation used for the relaxation rate in Sec. \ref{sec:exp} and \ref{sec:NMR}. The third column expresses the total dephasing rate for the resonance, which is a sum of the NV electron spin dephasing rate, $\Gamma_{2}^{(\rm p)}$, and the relevant damping rate of the target system: the dephasing rate of the P1 electron spin, $\Gamma_{2}^{(\rm t)}$, the longitudinal relaxation rate of the P1 electron spin, $\Gamma_{1}^{(\rm t)}$, or the dephasing rate of the P1 nuclear spin, $\Gamma_{2}^{'(\rm t)}$. In the fourth column, the Zeeman shift $\omega_e=-\tilde{\gamma}_eB$ is evaluated at the field where the transitions occur; the prime for the hyperfine-enhanced NMR transitions reminds that it is different than for the double-quantum EPR transitions, namely $\omega'_e\approx 2\omega_e$.} 	
\label{tab:P1strength}	
\end{table*}

The direct NMR transition corresponds to the situation where the NV electron spin interacts directly with a target nuclear spin, as was investigated in Sec. \ref{sec:NMR}, regardless of any hyperfine interaction with a nearby electron spin. For the P1 centre, this direct interaction is negligible in comparison with the hyperfine-enhanced interaction mediated by the P1 electron spin. Precisely, the ratio of the induced decay rates is 
\begin{equation}
\frac{\Gamma_{1,\rm res}^{\rm NMR,hyp}}{\Gamma_{1,\rm res\pm}^{\rm NMR}} \sim \left(\frac{A_\perp}{\omega_e}\right)^2 \left(\frac{\tilde{\gamma}_e}{\tilde{\gamma}_N}\right)^2.
\end{equation}
At the magnetic fields where these transitions occur ($B\approx 1000$ G), this fraction is $\approx 10^{5}$. This electron-mediated enhancement is significant and potentially paves the way for vastly improved sensing of nuclear spins through reporter electron spins.

\subsubsection{On the suppressed hyperfine-enhanced NMR transitions} \label{sec:NMRsuppressed}

Now we present an analysis of a pair of the hyperfine-enhanced NMR transitions in order to explain the different decay strengths seen and why only half of the transitions are detected in Sec. \ref{sec:exp}.

The P1 hyperfine interaction leads to transitions within the P1 centre between the states $\vert +\tfrac{1}{2},0\rangle\leftrightarrow \vert -\tfrac{1}{2},+1\rangle$ and $\vert +\tfrac{1}{2},-1\rangle\leftrightarrow \vert -\tfrac{1}{2},0\rangle$. The hyperfine-enhanced NMR transitions are achieved via a double-transition within the NV-P1 system involving one dipole-dipole interaction between NV and P1 electron and one hyperfine interaction between P1 electron and P1 nuclear spin leaving the P1 electron unchanged while both the NV and P1 nuclear spin are flipped. Consider the following NMR transitions within the NV-P1 system:
\begin{align*}
\left\vert0, +\tfrac{1}{2},+1\right\rangle&\rightarrow\left\vert-1, +\tfrac{1}{2},0\right\rangle\\
\left\vert0, +\tfrac{1}{2},0\right\rangle&\rightarrow\left\vert-1, +\tfrac{1}{2},+1\right\rangle.
\end{align*}
If we write out the full double transition via an intermediate state along with each type of transition we have
\begin{align*}
\left\vert0, +\tfrac{1}{2},+1\right\rangle&\xrightarrow{Dipole}\left\vert-1, -\tfrac{1}{2},+1\right\rangle\xrightarrow{Hyperfine}\left\vert-1, +\tfrac{1}{2},0\right\rangle\\
\left\vert0, +\tfrac{1}{2},0\right\rangle&\xrightarrow{Hyperfine}\left\vert0, -\tfrac{1}{2},+1\right\rangle\xrightarrow{Dipole}\left\vert-1, +\tfrac{1}{2},+1\right\rangle.
\end{align*}
The first of these transitions has a dipole-dipole transition of the form $\vert0, +\frac{1}{2},m_I^{(t)}\rangle\rightarrow\vert-1, -\frac{1}{2},m_I^{(t)}\rangle$ which has a spatial dependence of $\left(\frac{3\sin^2{\theta}}{r^3}\right)^2$ while the second has a dipole-dipole transition of the form $\vert0, -\frac{1}{2},m_I^{(t)}\rangle\rightarrow\vert-1, +\frac{1}{2},m_I^{(t)}\rangle$ which has a spatial dependence of $\left(\frac{3\sin^2{\theta}-2}{r^3}\right)^2$.

The total decay rate is the linear sum of all the contributing decays of atoms in the bath. Hence integrating these functions across all space gives the comparative strength of the transitions. Doing this in spherical coordinates gives
\begin{align*}
\int_0^{2\pi}\int_0^{\pi}\int_{r_{\rm min}}^{\infty}\left(\frac{3\sin^2{\theta}}{r^3}\right)^2 r^2 \sin{\theta} {\rm d}r {\rm d}\theta {\rm d}\phi = \frac{32\pi}{5 r_{\rm min}^3}\\
\int_0^{2\pi}\int_0^{\pi}\int_{r_{\rm min}}^{\infty}\left(\frac{3\sin^2{\theta}-2}{r^3}\right)^2 r^2 \sin{\theta} {\rm d}r {\rm d}\theta {\rm d}\phi = \frac{16\pi}{15 r_{\rm min}^3}.
\end{align*}
As a result, those transitions depending on $3\sin^2\theta$ are expected to be on average a factor of $\frac{32\pi}{5 r_{\rm min}^3}/\frac{16\pi}{15 r_{\rm min}^3} = 6$ times stronger than those transition that depend on $3\sin^2\theta-2$. The transitions depending on $3\sin^2\theta-2$ were not resolved in our NMR measurements in Sec. \ref{sec:exp} because of this expected weaker transition strength. This analysis assumes an ensemble average of bath spin positions and for our single NV case it will depend on the exact position of bath spins. 

\subsubsection{Simulation of the $T_1$-NMR spectrum} \label{sec:methods:H1simu}

In this section we briefly outline the method for numerically simulating the nuclear spin spectrum in Sec. \ref{sec:NMR}. The Hamiltonian of the NV centre is the same as in Eq. (\ref{eq:NVHam2}) and the nuclear spin Hamiltonian is the same as in Eq. (\ref{eq:TargHam}) with the correct nuclear gyromagnetic ratio replacing $\gamma_{\rm t}$. The interaction Hamiltonian is the dipole-dipole Hamiltonian (Eq. (\ref{eq:DDHam})). The simulation is done via evolution under the Lindblad equation from Eq. (\ref{eq:Lindblad}).

The superoperator formalism is used to allow timesteps longer than the dephasing time of both the NV and the target nuclear spin. In addition a background $T_1$ process was applied to the NV via taking timesteps of $\frac{T_1}{100}$ and decaying the diagonal elements of the density matrix with each timestep.

The initial state of the NV was taken to be $\vert m_S^{(\rm p)},m_I^{(\rm p)}\rangle = \vert 0,+1\rangle$ due to the nuclear spin polarisation near the GSLAC. The decay rate, $\Gamma_{1,\rm res}$, was found by fitting the population in $\vert 0,+1\rangle$ after evolution time t, to the function in Eq. (\ref{eq:PLfunction}).  

\subsubsection{Sensitivity} \label{sec:methods:sensitivity}

In this section we estimate the sensitivity of the method by comparing the signal caused by a target spin to the measurement noise. The measurement sequence consists of a 3-$\mu$s laser pulse followed by a wait time $\tau$ assumed to be much longer than 3 $\mu$s. The useful signal $I_s(\tau)$ is obtained by counting the photons within a read-out time $t_{\rm ro}=300$ ns. As a result, the PL signal is acquired only for a fraction $t_{\rm ro}/\tau$ of the total experiment time. Using Eq. (\ref{eq:PLfunction}), the total number of photons detected can be expressed as
\begin{equation}
{\cal N}(\Gamma_{1,\rm res},\tau)={\cal R}T_{\rm tot}\frac{t_{\rm ro}}{\tau}\left[1-{\cal C}+{\cal C}e^{-\Gamma_{1,\rm ph}\tau}\left(\frac{1}{4}+\frac{3}{4}e^{-\Gamma_{1,\rm res}\tau}\right)\right]
\end{equation}
where ${\cal R}$ is the photon count rate under continuous laser excitation and $T_{\rm tot}$ is the total acquisition time of the measurement. The change in the number of photons caused by the presence of a target spin inducing a relaxation rate $\Gamma_{1,\rm res}$ is 
\begin{eqnarray}
\Delta{\cal N}_{\rm signal}(\tau) & = & {\cal N}(0,\tau)-{\cal N}(\Gamma_{1,\rm res},\tau) \nonumber \\
& = & \frac{3{\cal R}T_{\rm tot}t_{\rm ro}{\cal C}}{4\tau}e^{-\Gamma_{1,\rm ph}\tau}\left(1-e^{-\Gamma_{1,\rm res}\tau}\right).
\end{eqnarray}
The photon shot noise associated with the measurement is
\begin{eqnarray}
\Delta{\cal N}_{\rm noise}(\tau) & = & \sqrt{{\cal N}(\Gamma_{1,\rm res},\tau)} \nonumber \\
& \approx & \sqrt{\frac{{\cal R}T_{\rm tot}t_{\rm ro}}{\tau}}
\end{eqnarray}
where we used the approximation ${\cal C}\ll 1$. The signal-to-noise ratio is then
\begin{eqnarray}
{\rm SNR}(\tau) & = & \frac{\Delta{\cal N}_{\rm signal}(\tau)}{\Delta{\cal N}_{\rm noise}(\tau)} \nonumber \\
& \approx & \sqrt{\frac{{\cal R}t_{\rm ro}T_{\rm tot}}{\tau}}\frac{3{\cal C}}{4}e^{-\Gamma_{1,\rm ph}\tau}\left(1-e^{-\Gamma_{1,\rm res}\tau}\right)
\end{eqnarray}
which corresponds to Eq. (\ref{eq:SNR}). \\

\section*{Acknowledgements}

This work was supported in part by the Australian Research Council (ARC) under the Centre of Excellence scheme (project No. CE110001027). L.C.L.H. acknowledges the support of an ARC Laureate Fellowship (project No. FL130100119).


\begin{thebibliography}{50}
\bibitem{NoteGamma}
Throughout this paper we use reduced gyromagnetic ratios $\tilde{\gamma}_i=\gamma_i/2\pi$ while the notation $\omega_i$ refers to an actual frequency rather than an angular frequency.
\end{thebibliography}
\end{document}